\documentclass[aps,prb,twocolumn,longbibliography]{revtex4-1}
\usepackage[english]{babel}
\usepackage[dvips]{graphicx}
\usepackage{color}
\usepackage{latexsym}
\usepackage{amsmath}
\usepackage{amsthm}
\usepackage{amsfonts}
\usepackage{amssymb}
\usepackage{romannum}
\usepackage{mathbbol}
\usepackage[colorlinks=true,citecolor=blue,linkcolor=blue]{hyperref}

\newcommand{\eps}{\varepsilon}

\newcommand{\bs}[1]{\boldsymbol{#1}}
\newcommand{\idxt}{t}
\newcommand{\idxb}{b}

\begin{document}

\pagenumbering{arabic}

\title{Transport through vertical graphene contacts under intense laser fields}
%
%
\author{P. Stadler}
\affiliation{Department of Microtechnology and Nanoscience (MC2), \\ Chalmers University of Technology, S-412 96 G{\"o}teborg, Sweden}

\author{T. L{\"o}fwander}
\affiliation{Department of Microtechnology and Nanoscience (MC2), \\ Chalmers University of Technology, S-412 96 G{\"o}teborg, Sweden}
\author{M. Fogelstr{\"o}m}
\affiliation{Department of Microtechnology and Nanoscience (MC2), \\ Chalmers University of Technology, S-412 96 G{\"o}teborg, Sweden}
\begin{abstract}
We theoretically study the electronic and transport properties of two graphene layers vertically coupled by an insulating layer under the influence of a time-periodic external light field. The non-adiabatic driving induces excitations of electrons and a redistribution of the occupied states which is manifested in the opening of gaps in the quasienergy spectrum of graphene. When a voltage is applied between the top and bottom graphene layers, the photo-induced nonequilibrium occupation modifies the transport properties of the contact.  We investigate the electronic and transport properties of the contact by using the nonequilibrium Green's function formalism. To illustrate the behavior of the differential conductance of the vertical contact under the light illumination, we consider two cases. First, we assume that both the bottom and top layers consist of graphene and second we consider a finite mass term in the bottom layer. We obtain that the differential conductance is strongly suppressed due to the opening of gaps in the quasienergy spectrum in graphene. Additionally, the conductance shows features corresponding to the tunneling of photoexcited electrons at energies of the van Hove singularity for both the top and bottom layers. 
In the case of a finite mass term in the bottom layer, the differential conductance can be directly related to the tunneling of photoexcited electrons.

\end{abstract}
%
%
%
%
%
%
%
%
%
%
%
\date{\today}
\maketitle
%
%
%

%
%
%
%
\section{Introduction}
\label{sec:intro}
The outstanding mechanical, optical and electronic properties of graphene make it an attractive material for next-generation technology~\cite{Ferrari:2015js}. Prominent examples are applications of graphene in optoelectronic devices such as photodetectors~\cite{mueller:2010,withers:2013,Liu:2014ku} and sensors~\cite{schedin:2010,Lee:2012,Pearce:2013ef}. The core of the improved functionality in graphene-based optoelectronic devices lies in the interaction of light and matter in low dimensions.  
Although a remarkable absorption of 2.3\% of incident light in monolayer graphene~\cite{nair:2008} is too weak to realize high-performance devices, several methods have been proposed to tune and enhance light-matter interaction in graphene. Among them, the absorption of graphene is enhanced by exciting surface plasmonic resonances\cite{ju:2011,thongrattanasiri:2012,grigorenko:2012,kim:2018} and by utilizing resonant structures to couple light with graphene \cite{ferreira:2012,hashemi:2013,zhao:2014}.

Apart from employing light-matter interaction in graphene to enhance the functionality of optoelectronic devices, graphene reveals several fundamental light-induced phenomena~\cite{glazov:2014}. A time-periodic external perturbation can turn a topologically trivial bandstructure into a topological one~\cite{lindner:2011,kitagawa:2010}. 
Depending on the polarization of the light field, gaps appear both at the Dirac points and other momenta in the Brillouin zone~\cite{syzranov:2008,lopez:2008,oka:2009,kibis:2010,calvo:2011,yang:2016,tenenbaum:2013,fruchart:2016,atteia:2017,Islam:2018,gu:2011}. The most prominent gap opens at energies  $\eps = \hbar\omega/2$ corresponding to a resonant absorption/emission of a photon with frequency $\omega$ between the valence and conduction band of graphene. Accompanied by the opening of a gap in bulk graphene, are chiral edge states. In particular, it has been discussed that nonlinear driving in graphene induces chiral edge photocurrents~\cite{Karch:2011} and the light-induced Hall effect without a magnetic field~\cite{oka:2009,mciver:2019}.

The periodic driving is in general studied within the Floquet formalism which is equivalent to Bloch theorem for periodic perturbation in time~\cite{shirley:1965}. 
Besides the opening of a gap, the temporal periodicity induces copies of the original
electronic bands spaced by integer multiples $n\omega$, the so-called
Floquet sidebands. Floquet bands have been observed by time- and angle-resolved photoemission spectroscopy of surface Dirac fermions in a topological insulator under circular light irradiation~\cite{wang:2013,mahmood:2016}. 
An alternative way to probe Floquet bands and the related opening of a gap in the band structure, is by conductance measurement. Signatures of  light-matter interaction in the conductance have been studied in planar two-terminal~\cite{syzranov:2008,oka:2009,gu:2011,calvo:2011,atteia:2017,Islam:2018} and multi-terminal contacts~\cite{kitagawa:2011,torres:2014,dehghani:2015}. The most dramatic effects of light-matter interation are manifested in a suppression of the conductance. 
In contrast to a time-periodic external light field, time-dependent modulation of gate or contact potentials and the accompanied photo-assisted tunneling of electrons lead to a great variety of interesting phenomena \cite{korniyenko:2016_1,korniyenko:2016_2,hammer:2013,pedersen:1998,platero:2004,kohler:2005}.

Motivated by the growing interest in nonequilibrium driving of graphene devices, we study a contact consisting of two graphene layers vertically separated by an insulating layer. A voltage is applied independently between the top and bottom layer allowing for tunneling between the layers. A similar setup consisting of a graphene/boron nitride/graphene heterostructure was discussed in Ref.~\onlinecite{mishchenko:2014} and it was demonstrated that resonant tunneling with the conservation of energy and momentum can be achieved by aligning the crystallographic orientation of the top and bottom graphene layers. In addition to the applied voltage, we assume that the top layer is driven to a nonequilibrium state by a time-periodic monochromatic light. 
We discuss two different scenarios. First, we assume that the vertical contact consists of two graphene layers. Second, we assume that a finite gap opens in the bottom layer. In our approach, we focus on noninteracting electrons in which Coulomb interaction is neglected. We study effects of both linear and circular polarization on the density of states, the occupation, and the differential conductance. 

The main results are the following. The differential conductance is strongly suppressed in a broad range of voltages. In a qualitative picture, the suppression of the differential conductance is related to opening of gaps in the quasienergy spectrum due to time-dependent driving. 
Various aspects of the suppression of the conductance due to light-matter interaction have been discussed in Refs.~[\onlinecite{syzranov:2008,oka:2009,gu:2011,calvo:2011,atteia:2017,Islam:2018,kitagawa:2011,torres:2014,dehghani:2015}]. Here, we extend these previous results of the differential conductance for vertical graphene contacts and the regime beyond the linear Dirac approximation. 

For strong driving and high frequencies, the momentum dependence of the optical matrix elements and the non-linearity of the graphene bandstructure becomes apparent in the differential conductance. Additional features occur at voltages that correspond to tunneling of photoexcited electrons to energies of the van Hove singularity in the top or bottom layer. For the case of a finite mass term in the bottom layer, the signatures in the differential conductance can be directly related to the tunneling of photo-excited electrons. 
Although we restrict the discussion to light absorption in graphene, the approach can be extended to other two-dimensional materials~\cite{wurstbauer:2017}.

The article is structured as follows. In Sec.~\ref{sec:model}, we introduce the continuous Hamiltonian and discuss the individual contribution of the Hamiltonian. From the continuous Hamiltonian, we derive a tight-binding Hamiltonian describing the light-matter interaction and the tunneling between the contacts. 
In Sec.~\ref{sec:method}  we describe the nonequilibrium Green's function formalism and the relations of the Green's functions to the physical properties. Sections~\ref{sec:dos}-\ref{sec:conductance} contain the results of our article. In Sec.~\ref{sec:dos} and~\ref{sec:occupation} we study the density of states and the occupation of graphene under light illumination, respectively. 
Finally, in Sec.~\ref{sec:conductance}, we discuss the differential conductance and show that the light-matter interaction induces suppression and enhancement of the conductance. In Sec.~\ref{sec:conclusions}, we summarize and conclude.

%
%
%
%
\section{Model}
\label{sec:model}
\begin{figure}[b!]			
	\includegraphics[width=0.8\linewidth,angle=0.]{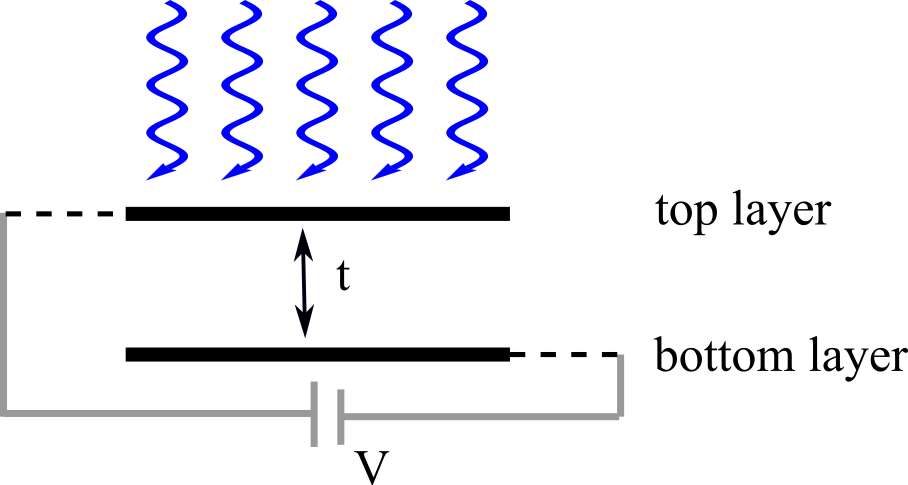} 
	\caption{Sketch of a contact with a top and bottom layer of graphene that are coupled by a tunneling element $t$. An electric field interacts with the top layer incident perpendicular to the layers. A voltage is applied between the top and the bottom layer.  }   
	\label{fig:fig1}
\end{figure}

We consider a vertical contact consisting of a top ($t$) and a bottom layer ($b$) which are coupled by an insulating layer described by a tunneling element $t$ as shown in Fig.~\ref{fig:fig1}. A voltage $V$ is applied between the top and the bottom layer and an external electric field perpendicular to the layers induces a nonequilibrium occupation of the electronic states. For simplicity, we assume that the electric field interacts only with the top layer and that the sample size is much smaller than the wavelength of the light such that we can regard the light field as being uniform in space.

The Hamiltonian of the vertical junction is given by
\begin{equation}
\hat{H}(t) = \hat{H}_{t}(t)+ \hat{H}_{b} + \hat{H}_{\mathrm{tun}} \, ,
\label{eq:Hamiltonian}
\end{equation}
with the Hamiltonians $\hat{H}_{t}(t)$ of the top layers, $\hat{H}_{b}$ of the bottom layer and the tunneling Hamiltonian $\hat{H}_{\mathrm{tun}}$. In the following we discuss the contributions to $\hat{H}(t)$ separately. We introduce a continuous time-dependent Hamiltonian $\hat{H}_t(t)$ of the top layer under light illumination and discretize it to obtain a tight-binding Hamiltonian.  Similar to the top layer, we then introduce the Hamiltonian of the bottom layer and the tunneling Hamiltonian.

\subsection{Hamiltonian of graphene under light illumination}
The Hamiltonian describing the motion of electrons with mass $m$ in the periodic lattice potential of graphene under light-radiation can be written as 
\begin{multline}
\hat{H}_{t}(t) = \int   d\bs{x} \, \hat{\Psi}_{t}^{\dagger} (\bs{x})  \left\{   \frac{[\hat{\bs{p}} -e\bs{A}(t)]^2}{2m}  
 \right. \\ \left. 
+ \sum_{\bs{x}^{\prime} =\bs{x}_A,\bs{x}_B }  V(\bs{x}-\bs{x}^{\prime})    \right\}  \hat{\Psi}_{t}(\bs{x}) \, ,
\label{eq:cH}
\end{multline}
with the field operators of the upper layer $\hat{\Psi}_{t}^{\dagger}(\bs{x})$ and  $\hat{\Psi}_{t}(\bs{x})$. The periodic lattice potential of graphene is composed of two triangular sublattices potentials $A$ and $B$, respectively. The light-matter coupling is obtained from the minimal substitution $\hat{\bs{p}} \rightarrow \hat{\bs{p}} -e\bs{A}(t)$ and the vector potential $\bs{A}(t)$ is modeled classically in the dipole approximation.  In the following studies we consider the case of linear polarized light  $\bs{A}(t)=(A_x\,\mbox{cos}(\omega t), 0, 0)$ and circlar polarized light  $\bs{A}(t)=(A_x\, \mbox{cos}(\omega t), A_y\, \mbox{sin}(\omega t), 0)$ with the frequency $\omega$ of the photons.

We then transform the continuous Hamiltonian $\hat{H}_{\idxt}(t)$ into a tight-binding Hamiltonian and expand the field operators in terms of the carbon $2 p_z$  wave functions $\Phi_{\idxt}(\bs{x}) $,
\begin{multline}
\hat{\Psi}_{\idxt}(\bs{x}) = \frac{1}{\sqrt{N}} \sum_{\bs{k}_{\idxt}, \bs{x}_A} e^{i \bs{k}_{\idxt}\bs{x}_A} \Phi_{\idxt}(\bs{x}-\bs{x}_A) \hat{a}_{\bs{k}_{\idxt}} 
\\ 
+
\frac{1}{\sqrt{N}} \sum_{\bs{k}_{\idxt}, \bs{x}_B} e^{i \bs{k}_{\idxt}\bs{x}_B} \Phi_{t}(\bs{x}-\bs{x}_B) \hat{b}_{\bs{k}_{\idxt}} \, ,
\label{eq:fieldoperators}
\end{multline}
with the annihilation operators $\hat{a}_{\bs{k}_{\idxt}}$ and  $\hat{b}_{\bs{k}_{\idxt}}$ of a particle with momentum $\bs{k}_{\idxt}$ on the sublattice $A$ and $B$ in the top layer, respectively.

\begin{figure}[t!]			
	\includegraphics[width=1\linewidth,angle=0.]{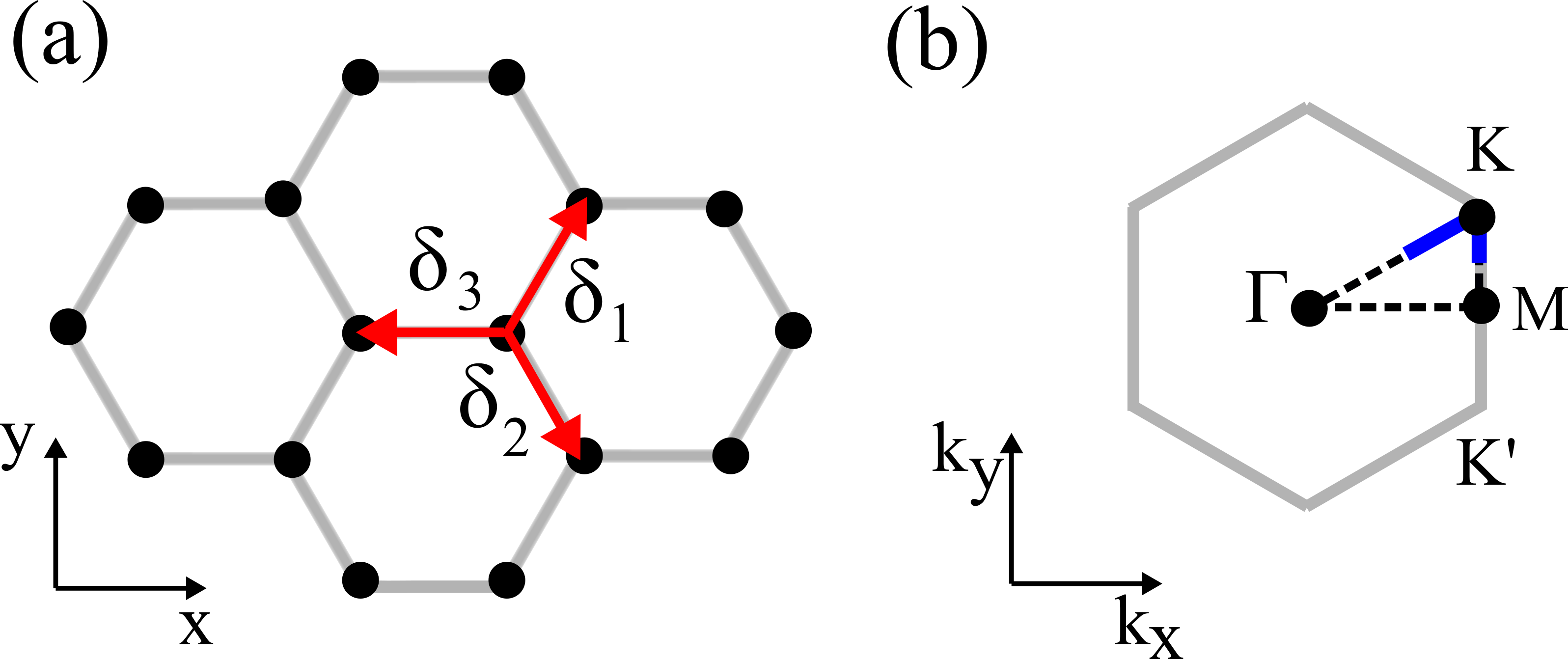} 
	\caption{(a) Lattice of graphene with the two sublattices $A$ (blue) and $B$ (green). The vectors $\bs{\delta}_i$ ($i=1,2,3$) connect the neighboring carbon atoms. (a) First Brillouin zone of graphene with the symmety points $\bs{\Gamma}$, $\bs{M}$, $\bs{K}$ and $\bs{K^{\prime}}$. The blue path  indicates the momentum dependence in the following figures.}   
	\label{fig:fig2}
\end{figure}

The tight-binding Hamiltonian is separated into a bare Hamiltonian $\hat{H}_0$ and a Hamiltonian $\hat{H}_I(t)$ descibing the interaction of graphene with the optical field,
$
\hat{H}_{\idxt}(t) = \hat{H}_0 + \hat{H}_I(t) \, .
$
It is convenient to transform the operators in sublatttice space to the space of the conduction and valence electrons. We write the creation operator of a quasiparticle in the conduction ($c$) and valence ($v$) band as $\hat{\psi}^\dagger_{\bs{k}_{\idxt}} = ({{\hat{a}{}}^{c}_{\bs{k}_{\idxt}}}^\dagger \,\, {{\hat{a}{}}^{v}_{\bs{k}_{\idxt}}}^\dagger)$. The bare Hamiltonian can then be written as
\begin{equation}
\hat{H}_0 = \sum_{\bs{k}_{\idxt}}  {{\hat{\psi}}_{\bs{k}_{\idxt}}}^{\dagger} \hat{\varepsilon}_{\bs{k}_{\idxt}}^{\phantom{\dagger}} 
{{\hat{\psi}}_{\bs{k}_{\idxt}}}^{\phantom{\dagger}}
\end{equation}
with the eigenenergies 
\begin{equation}
\hat{\varepsilon}_{\bs{k}_{\idxt}} = \begin{pmatrix}
\varepsilon_{\bs{k}_{\idxt}}^{\phantom{\lambda}}   & 0 \\
0 & -\varepsilon_{\bs{k}_{\idxt}}^{\phantom{\lambda}}  
\end{pmatrix} \, ,
\end{equation}
and $\varepsilon_{\bs{k}_{\idxt}} = \varepsilon^0_{\bs{k}_{\idxt}} -(\mu_{\idxt}-\eps_D)$ with $\varepsilon^0_{\bs{k}_{\idxt}} =  -\gamma_0 \vert f(\bs{k}_{\idxt})\vert$. The Dirac point energy $\eps_D$ sets the energy of the Dirac point relative to the chemical potential $\mu$. The interaction energy between the next-nearest neighbor carbon atoms is given by $\gamma_0$ and the chemical potential is $\mu_{\idxt}$.
In the above equation
\begin{equation}
f(\bs{k}_{\idxt}) = \sum_i \mathrm{e}^{i \bs{k}_{\idxt} \bs{\delta}_i }
\label{eq:fk}
\end{equation}
with the vectors $\bs{\delta}_i$ connecting neighboring carbon atoms. These vectors are given by $\bs{\delta}_1 = \frac{a}{2} ( 1 ,   \sqrt{3} )$, $\bs{\delta}_2= \frac{a}{2}(1 , \mathord-\sqrt{3} )$  and $\bs{\delta}_3 = ( -1 , 0 )$ with the nearest neighbor distance between the atoms $a = 1.42 \textrm{ \AA}$. The vectors $\bs{\delta}_i$ together with the lattice of graphene are shown in Fig.\ref{fig:fig2}(a).

Following Refs.~[\onlinecite{Stroucken:2011}] and [\onlinecite{Malic:2011}], we work in the Coulomb gauge $\nabla \cdot \bs{A} = 0$ and apply the dipole approximation since the momentum of the photon is negligible compared to the momentum of the electrons. The light-matter interaction can then be written as

\begin{equation}
\hat{H}_I(t) = \sum_{\bs{k}_{\idxt}}{{\hat{\psi}}_{\bs{k}_{\idxt}}}^{\dagger} \hat{M}_{\bs{k}_{\idxt}}^{\phantom{\dagger}} (t)
{{\hat{\psi}}_{\bs{k}_{\idxt}}}^{\phantom{\dagger}}
\label{eq:H0}
\end{equation}
with the light-matter interaction matrix
\begin{equation}
\hat{M}_{\bs{k}_{\idxt}}(t) = \begin{pmatrix}
M_{\bs{k}_{\idxt}}^{cc}(t)  & M_{\bs{k}_{\idxt}}^{cv}(t)\\
M_{\bs{k}_{\idxt}}^{vc}(t) &  M_{\bs{k}_{\idxt}}^{vv}(t) 
\end{pmatrix} \, ,
\label{eq:HI_final}
\end{equation}
and the elements 

\begin{align}
M_{\bs{k}}^{cc}(t) &= \mathrm{Re}[\bs{M}_{\bs{k}}]\bs{A}(t)  = -M_{\bs{k}}^{vv}(t)
\\
M_{\bs{k}}^{cv}(t) &= i \,\mathrm{Im}[\bs{M}_{\bs{k}}]\bs{A}(t)= M_{\bs{k}}^{vc^*}(t)   \, .
\end{align}

The time independent elements $\bs{M}_{\bs{k}}$ are
\begin{equation}
\bs{M}_{\bs{k}} = i \frac{e}{m}\hbar M e^{-i \phi_{\bs{k}}} \sum_i  e^{i\bs{k}\bs{\delta}_i} \frac{\bs{\delta}_i}{\vert \bs{\delta}_i\vert}  \, .
\label{eq:vecMK}
\end{equation}
with  $\phi_{\bs{k}} = \mathrm{arg} [f(\bs{k})] $ and a constant factor
$
M = \int d\bs{x} \, \Phi_u(\bs{x}) \partial_x \Phi_u(\bs{x}-
\vert \bs{\delta}_i \vert \hat{\bs{e}}_x )
$ 
with the unit vector in x-direction $\hat{\bs{e}}_x$.

The light-matter interaction can be divided into intraband coupling, $M_{\bs{k}_l}^{cc}(t)$ and $M_{\bs{k}_l}^{vv}(t) $, and interband coupling, $M_{\bs{k}_l}^{cv}(t)$ and $M_{\bs{k}_l}^{vc}(t) $. The intraband coupling describes processes in which an electron in the valence (conduction)- band absorbs/emits a photon with frequency $\omega$ and stays in the valence (conduction)-band. 
In a process of interband coupling, an electron scatters from the valence (conduction)-band to the conduction (valence) band and exchanges energy of the photon.

In the following, we combine all the constant values of Eq.~\eqref{eq:vecMK} in the amplitude of the vector potential $\bs{A}(t)$ and introduce the coupling strength 
\begin{equation}
\alpha_{x,y} = (e/m)\hbar M A_{x,y} \, .
\label{eq:couplingAlpha}
\end{equation}
in units of energy. 
To get an estimate of the magnitude of the vector potential, we consider the case of linear polarized light and convert the vector potential $\bs{A}(t)$ to an electric field by $\bs{E}(t) = \partial \bs{A}(t)/\partial t $. In Ref.~\onlinecite{mciver:2019}, a graphene sample was illuminated with an electric field amplitude $E_x = 4.0\times 10^7\textrm{ V/m}$\cite{mciver:2019} and a photon energy $\hbar\omega \approx 191 \textrm{ meV}$. Introducing the interaction energy of graphene $\gamma_0=2.6 \textrm{ eV}$ of the next-nearest neighbor carbon atoms, the photon energy corresponds to $ \hbar\omega \approx 0.07 \gamma_0$. The coupling strenght in Eq.~\eqref{eq:couplingAlpha} then is of the order $\alpha_{x,y} \approx 0.02 \gamma_0$ with the factor $M \simeq 3.0 \text{ nm}^{-1} $ \cite{Malic:2011}.

For the purpose of presentation, we show in following sections the results for slightly larger electric fields and photon energies. For example, in Secs.~\ref{sec:dos}-\ref{sec:conductance}, we assume an electric field amplitude of the order of $E_x = 0.1 \textrm{ V/\AA}$ and a photon energy $\hbar\omega \approx 1.7\textrm{ eV}$  corresponding to $\alpha_{x,y} \approx 0.1 \gamma_0$ and $\hbar\omega \approx 0.6\gamma_0$. We remark that high frequencies pulses with energies $\hbar\omega \approx 1.7\textrm{ eV}$ have been for instance applied in dielectrics to control the electric current by the electric field~\cite{schiffrin:2013}. Although in our approach, we assume a constant external field, high electric field amplitude can only be achieved by laser pulses. However, as pointed out in Ref.~\onlinecite{mciver:2019}, a constant field amplitude is still a valid approach if the envelope of the Gaussian pulse varies slowly in time~\cite{mciver:2019,sato:2019,novicenko:2017}. 
%
%

\subsection{Hamiltonian of the bottom lead and tunneling Hamiltonian}
The electrons can tunnel to the bottom layer which we model as graphene with a finite mass term. The Hamiltonian can be written as
\begin{equation}
\hat{H}_{\idxb}= \sum_{\bs{k}_{\idxb}}  {{\hat{\psi}}_{\bs{k}_{\idxb}}}^{\dagger} \hat{\varepsilon}_{\bs{k}_{\idxb}}^{\phantom{\dagger}} 
{{\hat{\psi}}_{\bs{k}_{\idxb}}}^{\phantom{\dagger}}
\label{eq:H0r}
\end{equation}
with the excitation energies
$
\varepsilon_{\bs{k}_{\idxb}} =\varepsilon^0_{\bs{k}_{\idxb}} -(\mu_{\idxb}-\eps_D)
$,
$
\varepsilon^0_{\bs{k}_{\idxb}} =- \gamma_0 \sqrt{\vert f(\bs{k}_{\idxb})\vert^2 + m_{\idxb}^2 }
$,
the chemical potential $\mu_{\idxb}$ and the mass $m_{\idxb}$. 
 
The tunneling Hamiltonian is given by 
\begin{equation}
\hat{H}_{\mathrm{tun}} = \sum_{\bs{k}_{\idxb}\bs{k}_{\idxt}} t_{\bs{k}_{\idxt}\bs{k}_{\idxb}} 
{{\hat{\psi}}_{\bs{k}_{\idxb}}}^{\dagger} {{\hat{\psi}}_{\bs{k}_{\idxt}}}^{\phantom{\dagger}} + \textrm{h.c.} \, ,
\end{equation}
with the tunneling elements $t_{\bs{k}_{\idxt}\bs{k}_{\idxb}} $. In the following we assume momentum conservation at tunneling and set $t_{\bs{k}\bs{k}^\prime} = \delta_{\bs{k}\bs{k}^\prime} t$.

\section{Method}
\label{sec:method}
Since we are interested in nonequilibrium transport properties, we apply the Keldysh Green's function approach to the Hamiltonian $\hat{H}$. In this method we express the physical properties in terms of Green's functions and calculate the corresponding Green's function.\cite{Cuevas-Scheer:2010,Rammer:2007}  To this end, we define the contour-ordered Green's function 
\begin{equation}
G^{\lambda\lambda^{\prime}}_{\bs{k}}(\tau,\tau^\prime) = -i 
\langle \mathcal{T}
\hat{a}^{\lambda}_{\bs{k}} (\tau)
\hat{a}_{\bs{k}}^{{\lambda^{\prime}\dagger}} (\tau^\prime) 
\rangle \, ,
\label{eq:GFdefinition}
\end{equation}
with the indices $\lambda=(c,v)$ and $\lambda^{\prime}=(c,v)$ indicating either the conduction band ($c$) or the valence band ($v$). The times $\tau$ and $\tau^{\prime}$ are located on the Keldysh contour and the contour-ordering operator is denoted as $\mathcal{T}$. 
To simplify the notation, we collect the Green's functions of the valence and conduction electrons in a matrix $\check{G}_{\bs{k}}(\tau,\tau^{\prime})$ defined by
\begin{equation}
\hat{G}_{\boldsymbol{k}}(\tau,\tau^{\prime})=
\begin{pmatrix}
G^{cc}_{\boldsymbol{k}}(\tau,\tau^\prime) 
& 
G^{cv}_{\boldsymbol{k}}(\tau,\tau^\prime)
\\
G^{vc}_{\boldsymbol{k}}(\tau,\tau^\prime) 
& 
G^{vv}_{\boldsymbol{k}}(\tau,\tau^\prime) 
\end{pmatrix} \, .
\label{eq:GFdefinition_matrix} 
\end{equation}

The general procedure then is to derive the Dyson equation for the Green's function in Eq.~\eqref{eq:GFdefinition_matrix}. This Green's function is given by
\begin{equation}
\hat{G}^{}_{\bs{k}}(\tau,\tau^{\prime}) = \hat{g}^{}_{\bs{k}}(\tau,\tau^{\prime})+\left(\hat{g}^{}_{\bs{k}} \circ \hat{M}_{\bs{k}} \circ \hat{G}^{}_{\bs{k}}\right) (\tau,\tau^{\prime}) \, .
\label{eq:leftDysonContour}
\end{equation}
with the convolution given by $(A\circ B)(\tau,\tau^\prime) = \int d\tau_1 A(\tau,\tau_1)B(\tau_1,\tau^\prime)$, the matrix  $\hat{M}_{\bs{k}}(\tau,\tau^\prime) = \hat{M}_{\bs{k}}(\tau)\delta (\tau-\tau^\prime)$ and the Green's function  
\begin{equation}
\hat{g}^{}_{\bs{k}}(\tau,\tau^{\prime}) = 
\begin{pmatrix}
{{g}_{\bs{k}}^{c}}^{}(\tau) & 0 \\
0 & {{g}_{\bs{k}}^v}^{}(\tau)
\end{pmatrix} \delta(\tau-\tau^\prime) \, ,
\label{eq:leftDysong0}
\end{equation}
corresponding to the Hamiltonian $\hat{H}_0$.
In the next step, the Dyson equation has to be transformed from the contour time to the real-time. \cite{Rammer:2007,Cuevas-Scheer:2010} After the transformation to the real time, every element in Eq.~\eqref{eq:GFdefinition_matrix} becomes a matrix in Keldysh space defined by
\begin{equation}
\check{G}^{}_{\bs{k}}(t,t^\prime) =
\begin{pmatrix}
\hat{G}^{11}_{\bs{k}}(t,t^\prime)
&
\hat{G}^{12}_{\bs{k}}(t,t^\prime)
\\[1mm]
\hat{G}^{21}_{\bs{k}}(t,t^\prime)
&
\hat{G}^{22}_{\bs{k}}(t,t^\prime)
\end{pmatrix} \, .
\label{eq:GFmatrix}
\end{equation}

In the above expression, the upper indexes $1$ or $2$ refer to the position of the times $t$ and $t^\prime$ on the Keldysh contour. A definition of all the Green's functions in Eq.~\eqref{eq:GFmatrix} and their relations is given in Appendix \ref{app:GFandSE}. From the elements in Eq.~\eqref{eq:GFmatrix} we can derive the retarded Green's function by
$
\hat{G}_{\boldsymbol{k}}^{R}(t,t^{\prime})=\hat{G}_{\boldsymbol{k}}^{11}(t,t^{\prime})-\hat{G}_{\boldsymbol{k}}^{12}(t,t^{\prime}) .
$
The advanced Green's function is related to the retarded Greens function by its hermitian conjugate 
$
\hat{G}^{{A}}_{\boldsymbol{k}}(t,t^{\prime}) = \hat{G}^{{R}^{\dagger}}_{\boldsymbol{k}}(t^{\prime},t) . 
$
The retarded ($R$), advanced ($A$) and the lesser ($12$) Greens functions constitute the building blocks by which we will express the charge current in the following section.

\subsection{Green's functions without light irradiation}
Following the procedure that we describe, we derive the Green's function for the bottom lead. Since we assume that the light only interacts with the upper layer, these Green's functions correspond to the Hamiltonian in Eq.~\eqref{eq:H0r} with momentum $\bs{k}_{\idxb}$. In Fourier space, the retarded Green's function is given by
$
{g}^{c,R}_{\bs{k}_{\idxb}}(\eps) = 1/(\eps+i\eta-\eps_{\bs{k}_{\idxb}} )
$
and 
$
{g}^{v,R}_{\bs{k}_{\idxb}}(\eps) = 1/(\eps+i\eta+\eps_{\bs{k}_{\idxb}} )
$, with a infinitesimal small real part $\eta$. The
lesser Green's function is 
$
\hat{g}^{12}_{\bs{k}_{\idxb}}(\eps) =-[\hat{g}^{R}_{\bs{k}_{\idxb}}(\eps) - \hat{g}^{A}_{\bs{k}_{\idxb}}(\eps) ] f_b(\eps)
$ 
with the temperature $T$ and the Fermi function $f_{b}(\eps)=\{1+\mathrm{exp}[(\eps-\mu_{\idxb})/k_B T]\}^{-1}$ with the chemical potential on the bottom layer $\mu_{\idxb}$.

\subsection{Green's functions under light irradiation}
\label{subsec:GFleftlead}
In this subsection, we derive the Dyson equation for the retarded and lesser Green's functions under light-irradiation. Because of the time-dependence in the optical field, we apply a discrete Fourier transformation to the Green's functions. To keep the notation simple, we replace the momentum $\bs{k}_t$ on the top layer with $\bs{k}$. 

From Eq.~\eqref{eq:leftDysonContour} and the relations between the Green's functions in Keldsyh space we can derive the Dyson equation for the retarded Green's function
\begin{equation}
\hat{G}^R_{\bs{k}}(t,t^{\prime}) = \hat{g}^R_{\bs{k}}(t,t^{\prime})+\left(\hat{g}^R_{\bs{k}} \circ \hat{M}_{\bs{k}} \circ \hat{G}^R_{\bs{k}}\right) (t,t^{\prime}) \, .
\label{eq:leftDyson}
\end{equation}
The Green's functions in Eq.~\eqref{eq:leftDyson} depends only on momentum $\bs{k}$  due to the dipole approximation in the light-matter interaction and the accompanied momentum conservation.
Similarly, we can derive the lesser Green's function 
\begin{equation}
\hat{G}^{12}_{\boldsymbol{k}}(t,t^{\prime})
=\left(\hat{G}^{R}_{\boldsymbol{k}}  
\circ 
{{\hat{g}{}}^{R}_{\boldsymbol{k}}}^{-1}  
\circ  
{\hat{g}{}}^{12}_{\boldsymbol{k}}
\circ  
{{\hat{g}{}}^{A}_{\boldsymbol{k}}}^{-1}  
\circ 
{\hat{G}{}}^{A}_{\boldsymbol{k}}  \right)(t,t^{\prime}) \, .
\label{eq:Gcc12}
\end{equation}

To compute the Green's functions  in Eqs.~\eqref{eq:leftDyson} and \eqref{eq:Gcc12} we have to apply a discrete Fourier transformation due to the time-dependence of the light-matter interaction. We can solve the Dyson equations by first transforming the time arguments to the Wigner representation with the relative time $\tau=t-t^{\prime}$ and the center-of-mass time $T= (t+t^\prime)/2$.\cite{Rammer:2007}  Second, we apply the Fourier transformation \cite{Cuevas-Scheer:2010}
\begin{equation}
\hat{G}^{R,12}_{\boldsymbol{k}}(t,t^{\prime}) = \sum_n \int \frac{d\varepsilon}{2\pi} \hat{G}^{R,12}_{\boldsymbol{k},n}(\varepsilon) e^{-i \varepsilon \tau} e^{-i n \omega T}
\end{equation}
and then numerically solve the Dyson equation by matrix inversion. To simplify later equations, we introduce the notation for the Fourier transformed Green's function
$
\hat{G}^{R,12}_{\boldsymbol{k},nm}(\varepsilon) =
\hat{G}^{R,12}_{\boldsymbol{k},n-m}(\varepsilon+(n+m)\omega/2)
$.
After the Fourier transformation of Eq.~\eqref{eq:leftDyson} we obtain
\begin{align}
\hat{G}^{R}_{\boldsymbol{k},nm}(\eps) = \hat{g}^{R}_{\boldsymbol{k},n}(\eps) \delta_{nm} 
-
& {\hat{g}^{R}_{\boldsymbol{k},n}} (\eps) \hat{M}^{+}_{\bs{k}} \hat{G}^{R}_{\boldsymbol{k},n+1m}(\eps)
\nonumber \\ -&
{\hat{g}^{R}_{\boldsymbol{k},n}}(\eps) \hat{M}^{-}_{\bs{k}} \hat{G}^{R}_{\boldsymbol{k},n-1m}(\eps)
\label{eq:GR_left}
\end{align}
with $\hat{g}^{R}_{\boldsymbol{k},n}(\eps) = \hat{g}^{R}_{\boldsymbol{k}}(\eps+n\omega)$ and the time-independent matrices 
\begin{equation}
\hat{M}_{\bs{k}_{}}^{\pm} = \begin{pmatrix}
\mathrm{Re}[\bs{M}_{\bs{k}}]\bs{A}^{\pm}  & i\mathrm{Im}[\bs{M}_{\bs{k}}]\bs{A}^{\pm}\\
-i\mathrm{Im}[\bs{M}_{\bs{k}}]\bs{A}^{\pm}&  -\mathrm{Re}[\bs{M}_{\bs{k}}]\bs{A}^{\pm}
\end{pmatrix} \, . 
\label{eq:HI_final}
\end{equation}
In Eq.~\eqref{eq:HI_final} $\bs{A}^{\pm}$ refers to the components of the vector potential $\bs{A}(t) = \bs{A}^{+} e^{i\omega t}+ \bs{A}^{-} e^{-i\omega t} $ which are proportional to positive ($+$) and negative ($-$) frequencies, respectively. For the case of linear polarization the components are $\bs{A}^{\pm} = (A_x/2,0,0)$ and for circular polarization $\bs{A}^{+} = (A_x/2,0,0)$ and $\bs{A}^{-} = (0,-i A_y/2,0)$. 
The unperturbed retarded Green's functions in Eq.~\eqref{eq:GR_left} of the conduction and valence electrons are given by
$
{g}^{c,R}_{\bs{k},n}(\eps) = (\eps+n\omega+i\eta-\varepsilon_{\bs{k}} )^{-1}
$
and 
$
{g}^{v,R}_{\bs{k},n}(\eps) = (\eps+n\omega+i\eta+\varepsilon_{\bs{k}} )^{-1}
$.

Applying the Fourier transform of the lesser Green's function in Eq.~\eqref{eq:Gcc12} gives
\begin{equation}
\hat{G}^{12}_{\boldsymbol{k},nm}(\eps)=\sum_{p}  \hat{G}^{R}_{\boldsymbol{k},np}(\eps)  \hat{g}^{R^{-1}}_{\boldsymbol{k},p}(\eps) \hat{g}^{12}_{\boldsymbol{k},p}(\eps) \hat{g}^{A^{-1}}_{\boldsymbol{k},p}(\eps) \hat{G}^{A}_{\boldsymbol{k},pm}(\eps)  \, .
\label{eq:G12_left}
\end{equation}

The unperturbed lesser Green's function is given by 
$
\hat{g}^{12}_{\bs{k},n}(\eps) =-[\hat{g}^{R}_{\bs{k},n}(\eps) - \hat{g}^{A}_{\bs{k},n}(\eps) ]f_{\idxt,n}(\eps)
$ 
with the temperature $T$ and the Fermi function $f_{\idxt,n}(\eps)=\{1+\mathrm{exp}[(\eps+n\omega-\mu_{\idxt})/k_B T]\}^{-1}$ with the chemical potential $\mu_{\idxt}$ on the top layer.

\subsection{Charge current}
With the Green's function derived in the previous sections, we can calculate the charge current. The current on the bottom lead can be expressed as
\begin{equation}
I_{\idxb} = -\frac{2e}{\hbar} \sum_{\bs{k}_{\idxb} \bs{k}_{\idxt}} \mathrm{Tr} \left\{ \tilde{G}^{12}_{\bs{k}_{\idxt} \bs{k}_{\idxb}}(t,t)t_{\bs{k}_{\idxb} \bs{k}_{\idxt}}-t_{\bs{k}_{\idxt} \bs{k}_{\idxb}}\tilde{G}^{12}_{\bs{k}_{\idxb} \bs{k}_{\idxt}}(t,t) \right\} \, ,
\label{eq:Irbare}
\end{equation}
with the trace over the Green's functions in the space of the conduction and valence electrons. In Eq.~\eqref{eq:Irbare} $\tilde{G}^{12}_{\bs{k}_{\idxt} \bs{k}_{\idxb}}(t,t)$ is the Green's functions corresponding to the full Hamiltonian $\hat{H}(t)$ taking the coupling between the leads into account to infinite order in the tunneling. However, to simplify the calculations, we restrict the discussion to the tunneling limit for which we can write the current as

\begin{equation}
I_{\idxb} = \frac{2e}{\hbar} \sum_n e^{in\omega T} I_{\idxb}^n \,
\label{eq:current}
\end{equation}
with
\begin{widetext}
\begin{align}
I_{\idxb}^n =
2t^2\sum_{\bs{k}_{\idxt}\bs{k}_{\idxb}}  \!\!\int  \frac{d\eps}{2 \pi} 
\mathrm{Re \, Tr}\!\left\{\hat{G}_{\bs{k}_{\idxt},n0}^{12}(\eps) \hat{g}_{\bs{k}_{\idxb},0}^{A}(\eps)
+
\hat{G}_{\bs{k}_{\idxt},n0}^{R}(\eps) \hat{g}_{\bs{k}_{\idxb},0}^{12}(\eps)\right\} \, .
\end{align}
\end{widetext}
with $t_{\bs{k}\bs{k}^\prime} = \delta_{\bs{k}\bs{k}^\prime} t$ and the momentum-independent tunneling element $t$. Since the summations over $\bs{k}_\idxt$ and $\bs{k}_\idxb$ separate in the tunneling limit, we can compute the summation over $\bs{k}_{\idxb}$ in the bottom layer analytically~\cite{hobson:1953,yuan:2010}. In Appendix~\ref{app:RecursiveGFSolution}, we present a recursive method to compute the Green's functions $\hat{G}_{\bs{k}_{},nm}^{R,A,12}$.

\section{Density of states under light irradiation}
\label{sec:dos}
In this section, we discuss the density of states and quasienergy spectrum of the graphene under light irradiation. We derive an effective Hamiltonian to study the splitting of quasi-eigenstates of graphene due to the light-matter interaction. The interaction of graphene with the light induces sidebands of the conduction and valence band, which are shifted by $\pm n \omega$ from the eigenenergies $\eps_{\bs{k}}$ of the bare graphene Hamiltonian. In the following, we restrict the discussion to a case of $n=0,\pm1,\pm2$ sidebands, since the effect of light on the quasienergy spectrum decreases with the number of sidebands. 

The quasispectrum of graphene under light irradiation has been discussed in literature using Floquet theory~\cite{shirley:1965,sambe:1973,syzranov:2008,oka:2009,gu:2011,calvo:2011,atteia:2017,Islam:2018,piskunow:2014,piskunow:2015,usaj:2014}.
For completness and discussions in later sections, we present here the results of the density of states of the band structure for momenta in the first Brillouin zone using the approach discussed in Sec.~\ref{sec:method}.

 \begin{figure}[t!]			
	\includegraphics[width=0.85\linewidth,angle=0.]{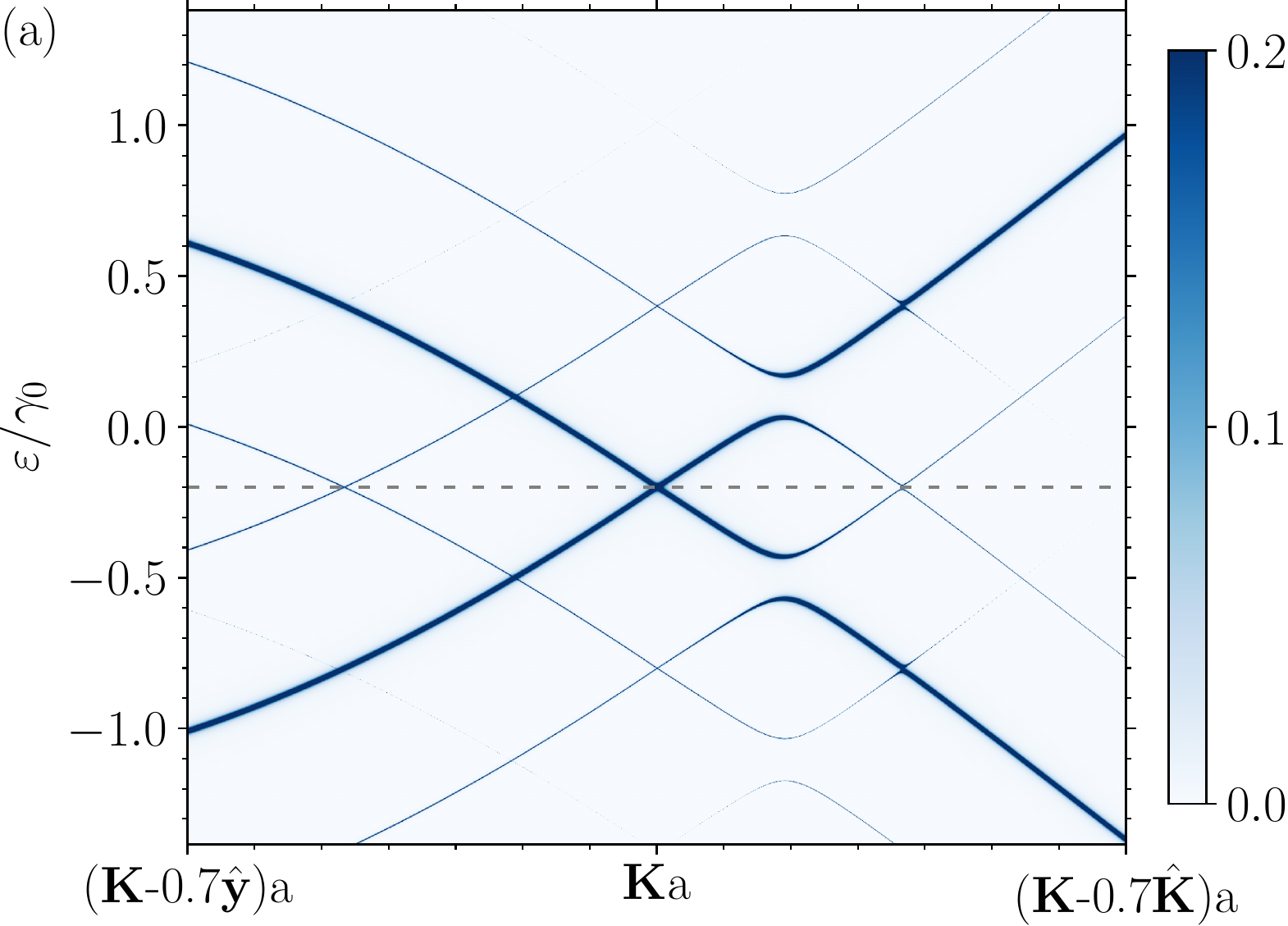} 
	\\ \vspace{5pt}
	\includegraphics[width=0.85\linewidth,angle=0.]{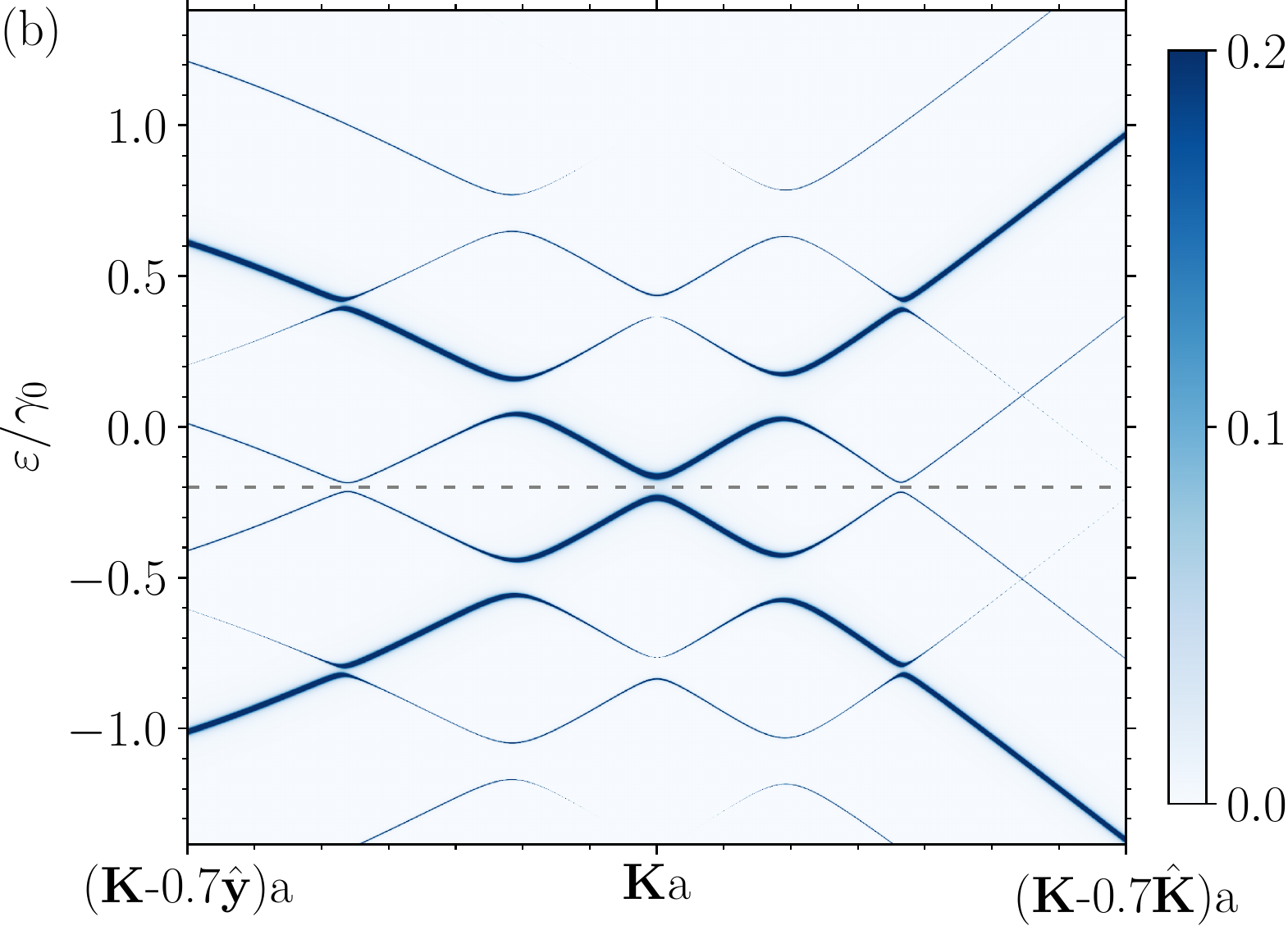} 
	\caption{Density of states $\rho_{\bs{k}}(\eps)$ in Eq.~\eqref{eq:dos} as a function of energy and momentum along the path $(\bs{K}-0.7 \hat{\bs{y}})\rightarrow  \bs{K} \rightarrow (\bs{K}-0.7 \hat{\bs{K}}) $ in the first Brillouin zone shown in Fig.~\ref{fig:fig2}(b). The unit vectors  in $\bs{y}$ and $\bs{K}$-direction are written as $\hat{\bs{y}}$ and $\hat{\bs{K}}$, respectively. The dashed line shows that Dirac point energy which is set to $\eps_D=-0.2\gamma_0$. The frequency is $\omega = 0.6\gamma_0$, $n=0,\pm1,\pm2$ and $\eta=10^{-4}\gamma_0$. In (a) we consider linear polarized light with $\bs{A}(t) = (A_x\mathrm{cos}(\omega t), 0 ,0)$ and $\alpha_x = 0.1 \gamma_0$. In this case the element $\Delta_{\bs{k}}$ vanishes along the path $(\bs{K}-0.7 \hat{\bs{y}})\rightarrow  \bs{K}$ and the spectrum remains ungapped.  In (b), we consider circular polarized light with $\bs{A}(t) = (A_x\mathrm{cos}(\omega t), A_y\mathrm{sin}(\omega t) ,0)$ and $\alpha_x = \alpha_y = 0.1 \gamma_0$. The density of state shows gaps at finite energies and becomes fully gapped at the Dirac point energy. The splitting close to the energies $\eps_{\bs{k}} = \omega/2+\eps_D$ is given by $\Delta_{\bs{k}}$.  }   
	\label{fig:fig3}
\end{figure}

The momentum-dependent density of states $\rho_{\bs{k}}(\eps) $ of graphene under light irradiation is given by
\begin{equation}
\rho_{\bs{k}}(\eps) = -\frac{1}{2\pi}\mathrm{Im\,Tr\,}\hat{G}^R_{\bs{k},00}(\eps) \, ,
\label{eq:dos}
\end{equation}
with the retarded Green's function in Eq.~\eqref{eq:GR_left}.  Figure~\ref{fig:fig3} shows the density of states of graphene along the path $(\bs{K}-0.7 \hat{\bs{y}})\rightarrow  \bs{K} \rightarrow (\bs{K}-0.7 \hat{\bs{K}})$ in the first Brillouin zone shown in Fig.~\ref{fig:fig2}(b). The unit vectors  in $\bs{y}$ and $\bs{K}$-direction are written as $\hat{\bs{y}}$ and $\hat{\bs{K}}$, respectively. 
We discuss both the case of linear polarized light [Fig.~\ref{fig:fig3}(a)] and circularly polarized light  [Fig.~\ref{fig:fig3}(b)].
Since a typical graphene sample is doped, we set the Dirac point energy to $\eps_D=-0.2\gamma_0$ such that states both above and below the Dirac point energy are occupied in an equilibrium state. The Dirac point energy in Fig.~\ref{fig:fig3} is indicated by the dashed line.

The interaction of light with graphene induces opening of gaps, the most prominent one is called the dynamical gap and occurs at energies $\eps_{\bs{k}} = \omega/2+\eps_D$. The transition corresponds to a resonant absorption/emission of a photon with energy $\hbar \omega$ between the conduction and valence band. Importantly, the matrix elements $\hat{M}_{\bs{k}}^{\pm}$ in Eq.~\eqref{eq:HI_final} depend on momentum and on the polarization of the light. For certain momenta and polarization, the matrix element vanishes and a gap opening is suppressed. In the case of circularly polarized light, the quasienergies become fully gapped at the Dirac point energy. We remark that the bandgap opening in graphene due to circularly polarized light infers a topological phase transition from a trivial to a topological band structure \cite{kitagawa:2010,kitagawa:2011,oka:2009,lindner:2011}.

\begin{figure}[t!]			
	\includegraphics[width=0.49\linewidth,angle=0.]{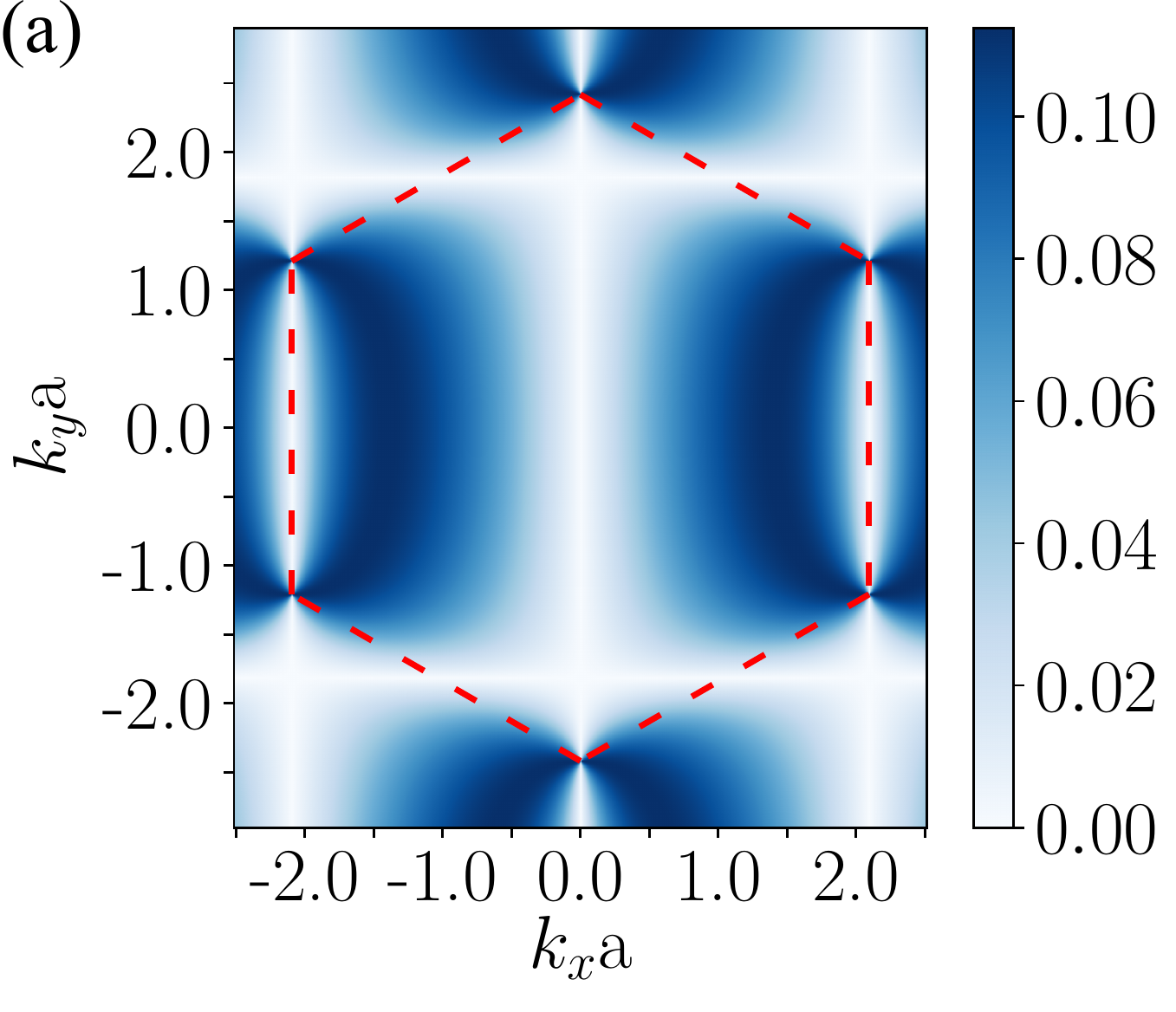} 
	\includegraphics[width=0.49\linewidth,angle=0.]{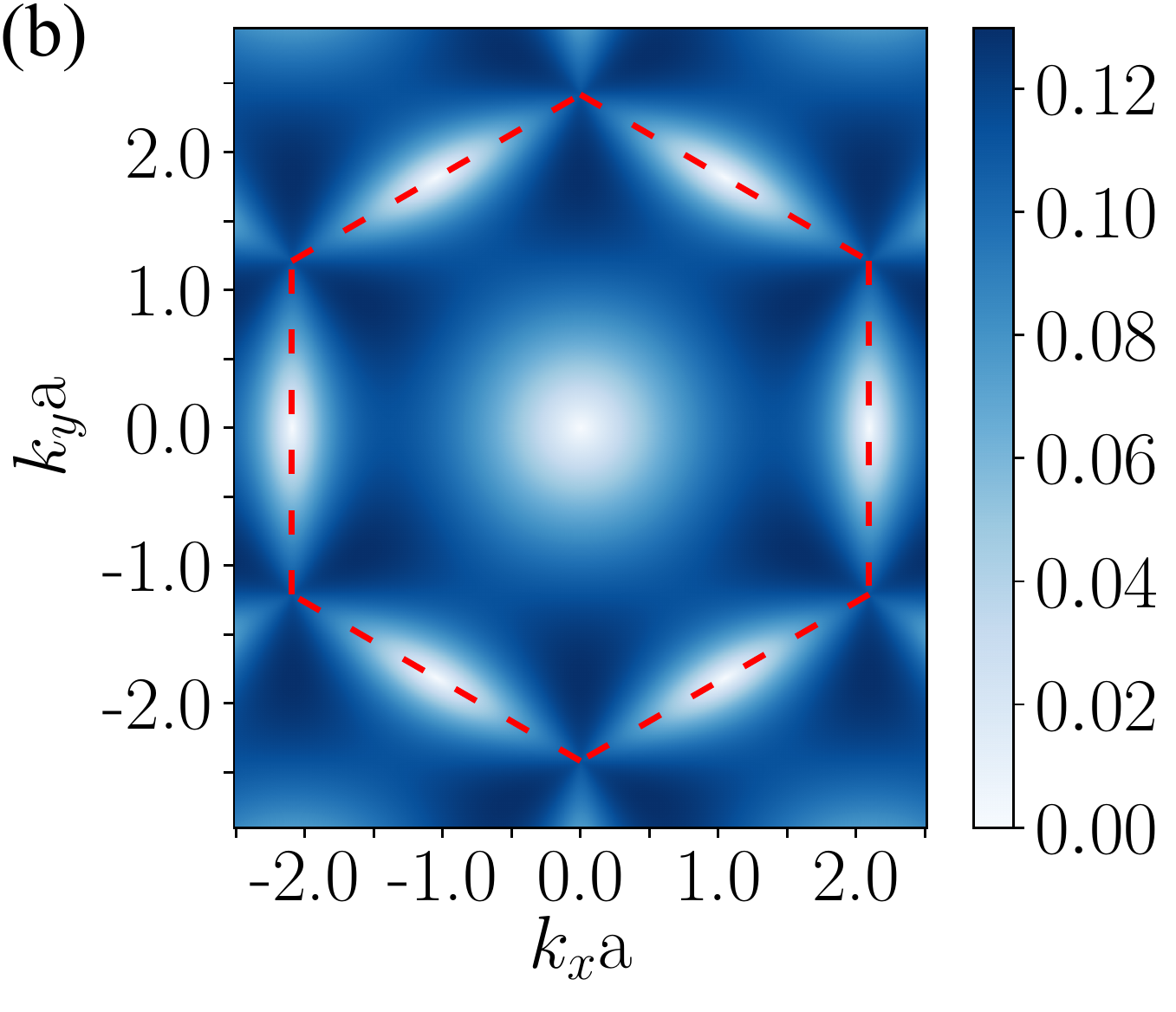} 
	\caption{Absolute value of the optical elements $\Delta_{\bs{k}} = \vert \mathrm{Im}[\bs{M}_{\bs{k}}]\bs{A}^{\pm} \vert$ for (a) linear polarized light and (b) circular polarized light. The parameters are the same as in Fig.~\ref{fig:fig3}. The red dashes line shows the boundary of the first Brillouin zone.  In (a), the element vanishes along the path $\bs{M} \rightarrow \bs{K}$ and the density states remains ungapped as shown in Fig.~\ref{fig:fig3}(a). In (b) the the optical matrix element $\Delta_{\bs{k}}$ is finite when approaching the $\bs{K}$ point and the density of states is fully gapped as shown in Fig.~\ref{fig:fig3}(b).  }   
	\label{fig:fig4}
\end{figure}

To further study the effect of the light-matter interaction on graphene, we restrict the discussion to a minimal model, in which we consider the emission or absorption of one photon such that $n=0,\pm 1$.  An effective Hamiltonian can be derived by comparing the retarded Green's function in Eq.~\eqref{eq:GR_left} with the definition of a retarded Green's function $\hat{G}^R_{\bs{k},nm}(\eps)=(\eps+i\eta - \hat{H}_{\bs{k},nm}^{\mathrm{eff}})^{-1}$. We can then write the effective Hamiltonian as 
\begin{equation}
\hat{H}_{\bs{k},nm}^{\mathrm{eff}} = 
\begin{pmatrix}
{\hat{\varepsilon}}_{\bs{k}} + \mathbb{1}\omega & -\hat{M}^+_{\bs{k}} & 0  
\\
-\hat{M}^{-}_{\bs{k}}  &   \hat{\varepsilon}_{\bs{k}}&  -\hat{M}^{+}_{\bs{k}} 
\\
0 &    -\hat{M}^{-}_{\bs{k}} &  \hat{\varepsilon}_{\bs{k}}-\mathbb{1}\omega
\end{pmatrix} \, ,
\label{eq:H_eff}
\end{equation}
Assuming that $\hat{M}^{\pm}_{\bs{k}}=0$, the eigenvalues of $\hat{H}_{\bs{k},nm}^{\mathrm{eff}} $ are given by $E^{c,v}=\pm\eps_{\bs{k}}$ and the sidebands $E^{c,v}_{+}=\pm \eps_{\bs{k}}+\omega$ and $E^{c,v}_{-}=\pm \eps_{\bs{k}}-\omega$ which are shifted by $\pm \omega$ from the original conduction and valence band, respectively.

To study the dynamical gap, we simplify the effective Hamiltonian and only consider the conduction band coupled with the first sideband of the valence band. Such an effective model corresponds to a $2x2$ matrix in $\hat{H}_{\bs{k}}^{\mathrm{eff}}$ and describes the gap at finite energies. The eigenvalues for such a system are given by
\begin{equation}
\label{eq:dynamicgap}
\eps^{\pm}_{\bs{k}} = \frac{\omega}{2} \pm \sqrt{\left(\eps_{\bs{k}}-\frac{\omega}{2}\right)^2+\vert \Delta_{\bs{k}} \vert^2 } \, .
\end{equation}
with the off-diagonal matrix element $\Delta_{\bs{k}} = \vert \mathrm{Im}[\bs{M}_{\bs{k}}]\bs{A}^{\pm} \vert$ being proportional to the absolute value of the imaginary part of the light-matter matrix element connecting the valence and conduction band. Close to energies $\varepsilon_{\bs{k}} \simeq \omega/2+\eps_D$, the splitting between the eigenvalues is therefore $2\Delta_{\bs{k}}$ and hence the gap is proportional to the coupling strength $\alpha_{x,y}$. The general behavior of gap openings under light illumination has been discussed for instance in Refs.~\onlinecite{atteia:2017} and \onlinecite{zhou:2011}.

Since the dynamical gap is proportional to $\Delta_{\bs{k}}$ we can understand the gap opening at finite energies in Fig.~\ref{fig:fig3} by consider the momentum dependence of $\Delta_{\bs{k}}$. 
Fig.~\ref{fig:fig4} shows the momentum dependence $\Delta_{\bs{k}}$ for (a) linear and (b) circular polarized light with the dashed line denoting the first Brillouin zone. When changing the momentum from $(\bs{K}-0.7 \hat{\bs{y}}) \rightarrow  \bs{K} $ [see Fig.~\ref{fig:fig3}(a)], the element $\Delta_{\bs{k}}$ stays zero and states will be ungapped along the  path $(\bs{K}-0.7 \hat{\bs{y}}) \rightarrow  \bs{K} $. However, the element $\Delta_{\bs{k}}$ is finite when changing the momentum along the path $\bs{K}\rightarrow (\bs{K}-0.7 \hat{\bs{K}})$ and the spectrum is gapped. In Fig.~\ref{fig:fig4}(b), the element $\Delta_{\bs{k}}$ is finite when approaching the $\bs{K}$ point from both $(\bs{K}-0.7 \hat{\bs{y}})$ and $(\bs{K}-0.7 \hat{\bs{K}})$, and the spectrum is gapped at energies $\eps_{\bs{k}}\simeq \omega/2+\eps_D$ close to the $\bs{K}$ point. 

\section{Electronic occupation}
\label{sec:occupation}
In this section, we discuss the modification of the occupation on the upper lead due to light irradiation. Direct measurement of the nonequilibrium occupation has been performed on the surface states of a topological insulator by time and angle-resolved photoemission spectroscopy~\cite{wang:2013}.

From the definition of the Green's functions, we can write the energy and $\bs{k}$-dependent occupation of the top layer as
\begin{equation}
n_{\bs{k}}(\eps) = -i \, \mathrm{Tr}\, \hat{G}^{12}_{\bs{k},00}(\eps) \, ,
\end{equation}
with the lesser Green's function given by Eq.~\eqref{eq:G12_left}. In a simple picture, the occupation is given by the density of states multiplied by the Fermi functions at different energies $f_t(\eps)$ and $f_t(\eps\pm \omega)$.

\begin{figure}[t!]			
	\includegraphics[width=0.9\linewidth,angle=0.]{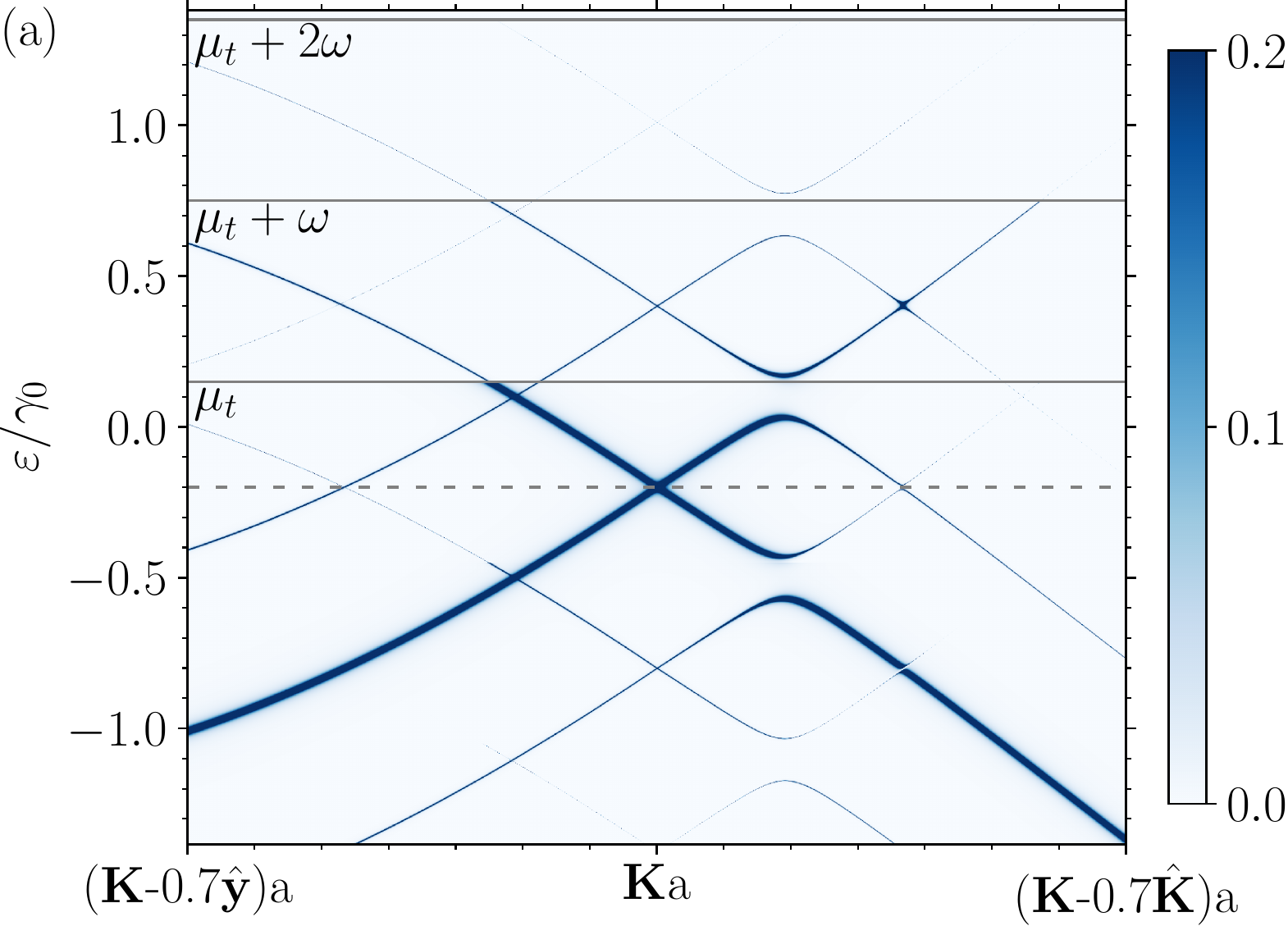} 
	\\ \vspace{5pt}
	\includegraphics[width=0.9\linewidth,angle=0.]{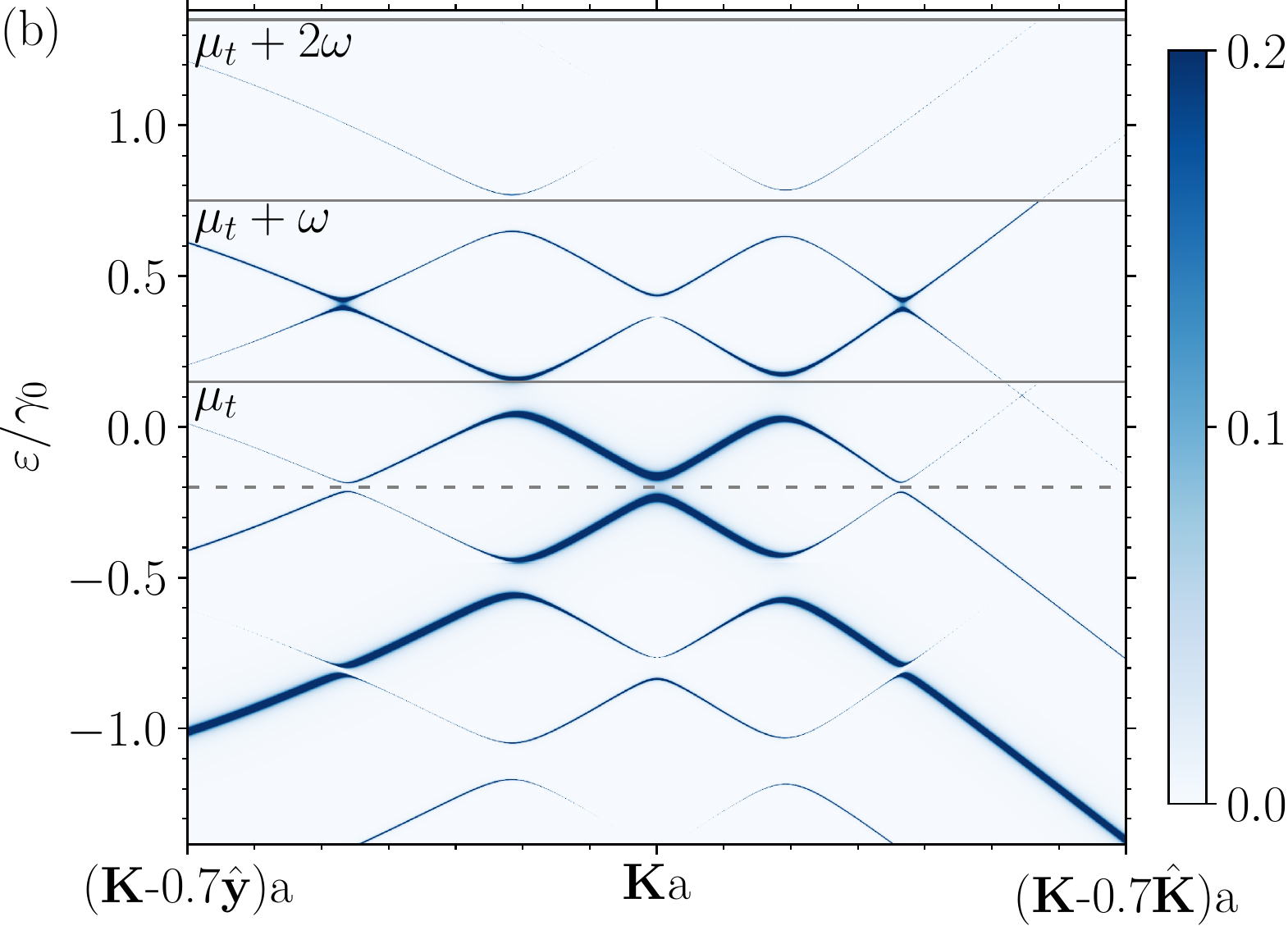} 
	\caption{Occupation of the top lead along the path $(\bs{K}-0.7 \hat{\bs{y}}) \rightarrow  \bs{K} \rightarrow (\bs{K}-0.7 \hat{\bs{K}}) $ in the first Brillouin zone and as a function of energy $\varepsilon$. The frequencies of the optical field is $\omega=0.6\gamma_0$ and the chemical potential is $\mu_{\idxt} = 0.15 \gamma_0$, $\eps_D=-0.2\gamma_0$, $n=0,\pm1,\pm2$, $T=0$ and $\eta=10^{-4}\gamma_0$. Without the light-matter interaction, state are occupied at zero temperature to the chemical potential $\mu_{\idxt}$. A finite optical field occupies states above the chemical potential with maximal energy $\mu_{\idxt}+2\omega$.  In (a), the light is circular polarized with $\alpha_{x}=\alpha_{y} = 0.1\gamma_0$ and in (b) the light is linear polarized in $x$-direction with $\alpha_{x}= 0.1\gamma_0$ and $\alpha_{y} =0$.  
	} 
	\label{fig:fig5}
\end{figure}

Figure~\ref{fig:fig5} shows the occupation along the path $(\bs{K}-0.7 \hat{\bs{y}}) \rightarrow  \bs{K} \rightarrow (\bs{K}-0.7 \hat{\bs{K}}) $ in the first Brillouin zone and as a function of energy $\eps$ for the same parameters as in Fig.~\ref{fig:fig4}. We consider the case of linear [Fig.~\ref{fig:fig5}(a)] and circular polarized light [Fig.~\ref{fig:fig5}(b)] with a frequency of $\omega=0.6\gamma_0$. We set the chemical potential to $\mu_t = 0.15\gamma_0$, the Dirac point energy to $\eps_D=-0.2\gamma_0$ and we consider $n=0,\pm1,\pm2 $ sidebands. 

\begin{figure*}[ht!]			
	\hfill
	\begin{center}
		\includegraphics[width=0.5\linewidth,angle=0.]{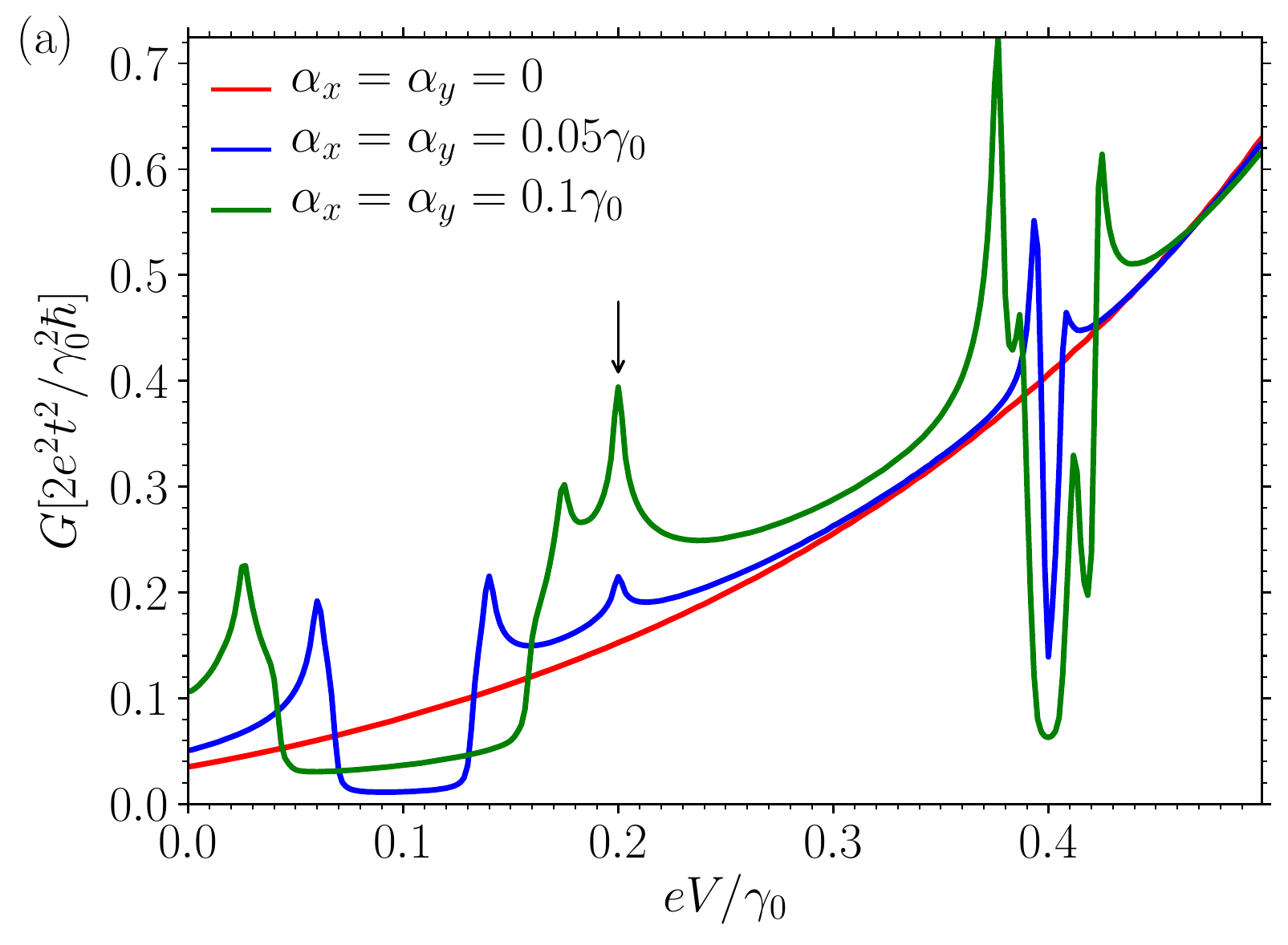} 
		\hspace{15pt}
		\raisebox{6mm}{\includegraphics[width=0.4\linewidth,angle=0.]{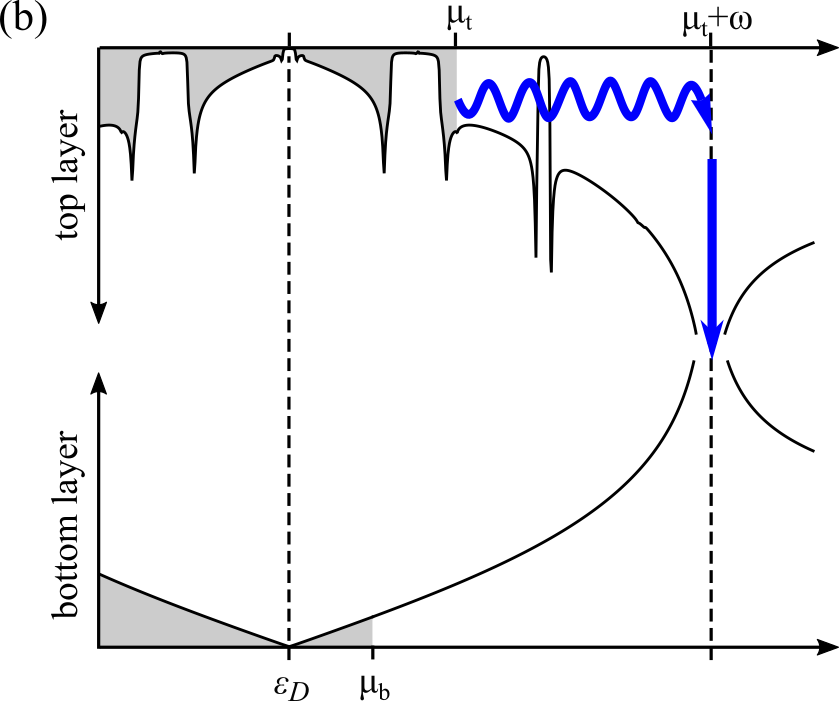}}
	\end{center}
	\caption{(a) Differential conductance for circularly polarized light and different coupling strength $\alpha_x$ and $\alpha_y$. The parameters are $\omega = 0.6 \gamma_0$, $T=0$, $n=0,\pm1,\pm2$, $\eta= 0.001\gamma_0$. The Dirac point energy is set to $\eps_D=-0.2\gamma_0$ such that the differential conductance is suppressed close to voltages $eV \simeq \omega/2+\eps_D = 0.1\gamma_0$. The light-matter interaction empties states and the differential conductance decreases. The finite conductance in the gap depends on the value of $\eta$. The steps in the case of $\alpha_x=\alpha_y=0.1\gamma_0$ close to $eV\approx 0.03\gamma_0$ and $eV\approx 0.17\gamma_0$ are a signature of the non-linear band structure of graphene for $\bs{k} \gg (\bs{K},\bs{K}^\prime)$ and the momentum dependence of the optical matrix elements. (b) Sketch of the density of states of the top and bottom layer showing the process of photoexcitation of an electron at energy $
	\mu_{\idxt}$. The photoexcited electron has energy matching the van Hove singularity of the top layer and tunnels in the van Hove singularity of the bottom layer. The process gives rise to a peak in (a) indicated by the black arrow.}   
	\label{fig:fig6}
\end{figure*}

Without light-matter interaction and at zero temperature, states can be occupied to the chemical potential $\mu_t$. A finite coupling strength excites electrons and occupies states above the Fermi level $\mu_t$. The highest accessible states at zero temperature are at energies $\eps=\mu_{\idxt}+2\omega$ corresponding to absorptions of two photons with energy $\omega$ by one electron.

It is interesting to note that the gaps in Fig.~\ref{fig:fig5}(b) close to energies $\eps \approx 0.1\gamma_0$ are different in size. These gaps correspond to the resonant absorption or emission of a single photon with energy $\omega$. The different gap sizes occur due to the matrix elements shown in Fig.~\ref{fig:fig4}(b) whose magnitude dependent on the direction from which the Dirac point is approached. The different sizes of the gap at finite energy vanish within the Dirac approximation of the band structure and the optical matrix elements.  The effect is a result of the consideration of the full momentum dependence of energies and optical matrix elements in the first Brillouin zone.

 \section{Differential conductance}
 \label{sec:conductance}
With the current derived in Eq.~\eqref{eq:current}, we can discuss different scenarios for transport through a graphene contact and identify signatures of the light-matter interaction in transport measurements. We focus on the differential conductance which can be written as
\begin{equation}
G(V) = \frac{dI^0_b(V)}{dV} \, ,
\end{equation}
with the dc-current given by Eq.~\eqref{eq:current}. Throughout the discussion, we assume that circularly polarized photons are absorbed in the top layer and the photo-excited electrons tunnel to the bottom layer.

Motivated by the fact that an underlying substrate~\cite{zhou:2007} or Coulomb interaction~\cite{Stroucken:2011} opens a gap in the band structure of graphene, we discuss two scenarios: (i) we assume that both the top and the bottom layer consist of graphene. (ii) We discuss the case that the top layer consists of graphene and the bottom layer consists of graphene with a finite mass term opening a gap in the band structure.

\begin{figure*}[t!]			
	\hfill
	\begin{center}
		\includegraphics[width=0.5\linewidth,angle=0.]{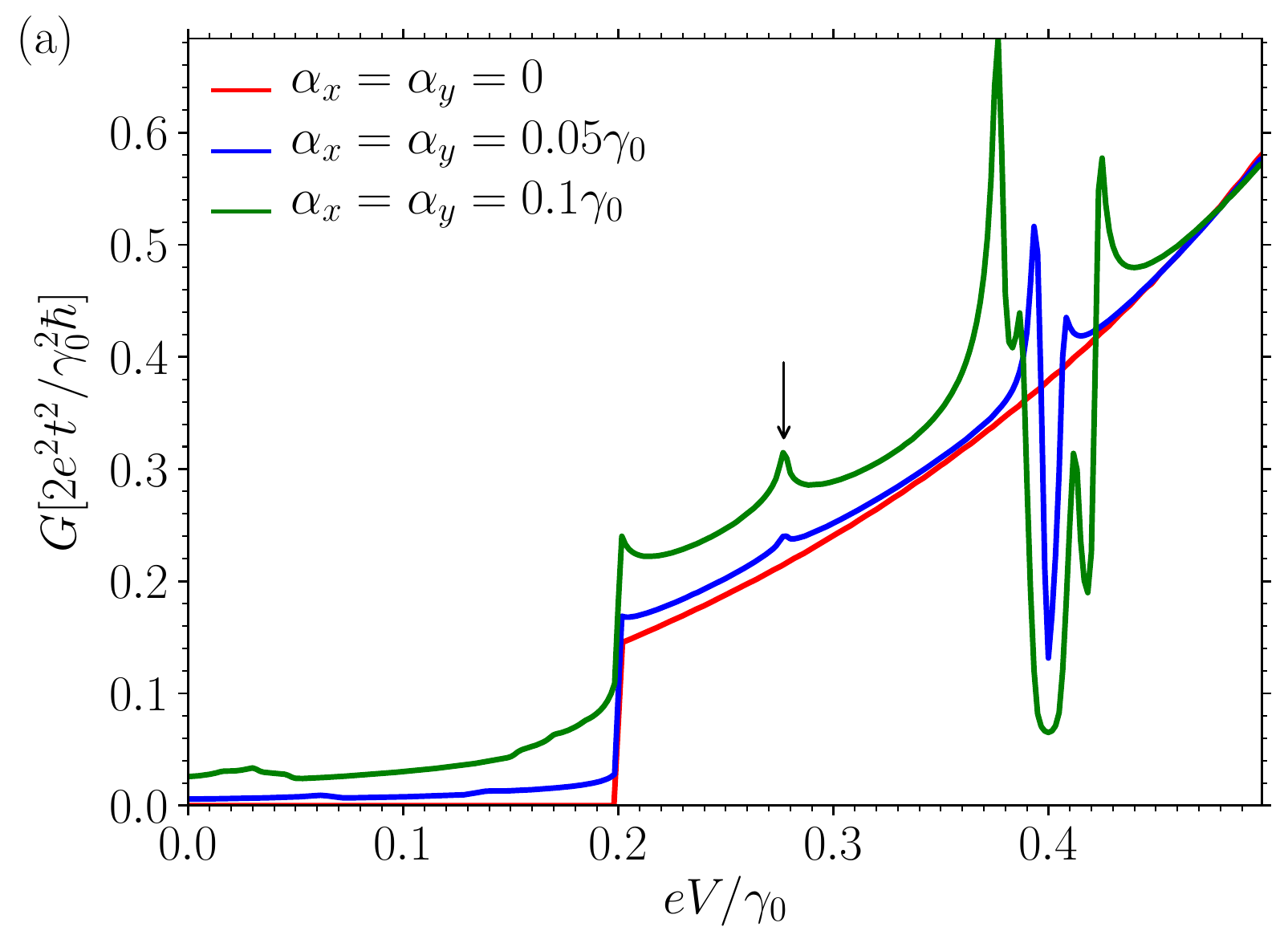} 
		\hspace{15pt}
		\raisebox{6mm}{\includegraphics[width=0.4\linewidth,angle=0.]{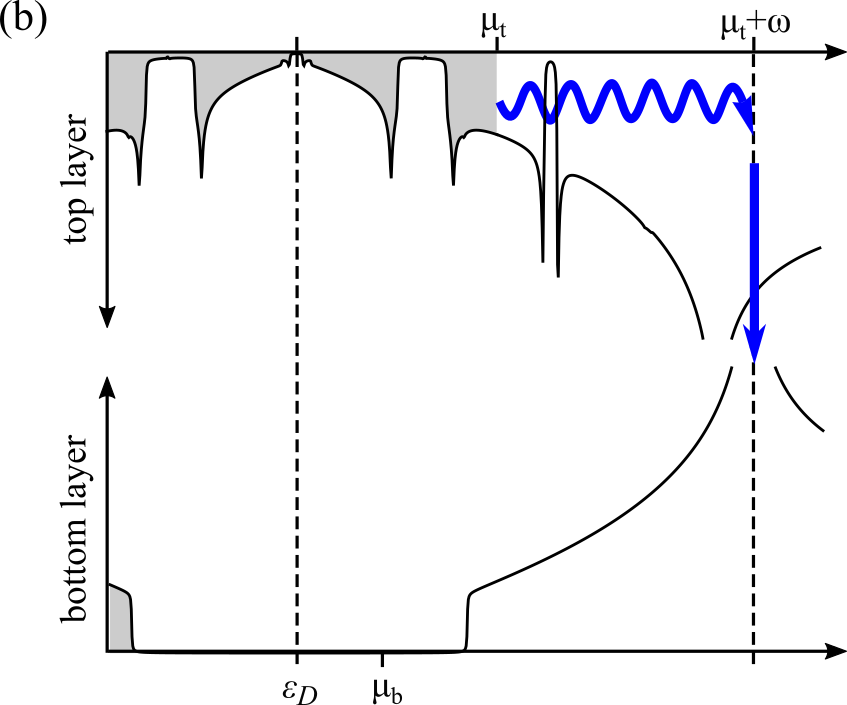}}
	\end{center}
	\caption{(a) Differential conductance for circularly polarized light and different coupling strength $\alpha_x$ and $\alpha_y$. The parameters are $\omega = 0.6 \gamma_0$, $\eps_D = -0.2\gamma_0$, $n=0,\pm1,\pm2$, $T=0$, $\eta= 0.001\gamma_0$. In comparison to Fig.~\ref{fig:fig6}(a), here we assume that the bottom layer consists of graphene with a finite mass term $m_b = 0.4 \gamma_0$ which opens a gap of $2m_b$. For $\alpha_x=\alpha_y=0$, the conductance jumps to a finite value when the voltage reaches the gap edge of the bottom graphene layer. Because of the Dirac point energy, the step-like increase sets in at $eV = 0.2\gamma_0$. A finite light-matter interaction excites electrons above the gap edge of the bottom lead and the conductance is finite even for voltages $eV< 0.2\gamma_0$. (b) Sketch of the density of states of the top and bottom layer showing the process of absorption of a photon at energy $\mu_{\idxt}$ and the accompanied tunneling in the van Hove singularity of the bottom layer. The process gives rise to the peak at $eV=0.28\gamma_0$ which is indicated in (a) by the black arrow. Because of the gap in the density of states of the bottom layer, the van Hove singularity is shifted to larger energies compared to the energy of the van Hove singularity of the top layer. }   
	\label{fig:fig7}
\end{figure*}

\subsection{Tunneling between graphene contacts}
Figure~\ref{fig:fig6}(a) shows the differential conductance for circularly polarized light and different values of the coupling strength $\alpha_{x,y}$.  
The frequency of the incoming light is set to $\omega = 0.6\gamma_0$ and the temperature is $T=0$. Since a typical graphene sample is doped we assume a finite Dirac point energy of  $\eps_D = -0.2\gamma_0$ relative to the chemical potential $\mu_t$.

We first consider the case without light-matter interaction ($\alpha_x=\alpha_y = 0$). At small voltages, $eV \ll \gamma_0$, the differential conductance increases with the applied voltage and depends nonlinearly on the voltage for $eV \rightarrow \gamma_0$. Since we consider the tunnelling current in Eq.~\eqref{eq:current} this behavior is reflected by the overlap between the density of states of the top and bottom layer. 
The density of states of single-layer graphene is proportional to $\eps$ for $\eps \ll \gamma_0 $ and approaches the van Hove singularity at $\eps =  0.8\gamma_0$ assuming finite doping $\eps_D = -0.2\gamma_0$ \cite{castro:2009}.

As discussed in the previous section, a finite coupling strength opens a gap in the quasienergy spectrum of graphene. Even for moderate coupling strength  $\alpha_x = \alpha_y = 0.05 \gamma_0$, the opening of a gap in the quasienergy spectrum is manifested in a strong suppression of the differential conductance at certain voltages. 

The first suppression of the current close to $eV=0.1\gamma_0$ corresponds to resonant absorption of a single photon between the conduction and valence band. Since the Dirac point energy is at $\eps = -0.2\gamma_0$ and the frequency of the light is $\omega = 0.6\gamma_0$, the suppression of the conductance occurs at $eV \approx \omega/2+\eps_D = 0.1\gamma_0$. The second strong suppression of the differential conductance occurs close to $eV=0.4\gamma_0$ and corresponds to resonant absorption of two photons between the conduction and valence band. 

At strong coupling $\alpha_x = \alpha_y = 0.1 \gamma_0$, the differential conductance is suppressed over a larger range of voltages compared to the moderate coupling strength $\alpha_x = \alpha_y = 0.05 \gamma_0$ since the gap in the quasienergy spectrum increases with increasing coupling strength. 
Interestingly, the suppression and the enhancement of the conductance close to voltages $eV \approx 0.1\gamma_0$ gradually decreases and increases at strong coupling $\alpha_x = \alpha_y = 0.1 \gamma_0$. The steps are a signature of the nonlinear band structure of graphene and the momentum dependence of the optical matrix elements for momenta $\bs{k} \gg (\bs{K},\bs{K}^\prime)$. 
The gradual increase and decrease can be understood by considering for instance the conductance at voltage $eV=0.15\gamma_0$ in Fig.~\ref{fig:fig6}(a).  Figure \ref{fig:fig5}(b) shows the occupation corresponding to the voltage $eV = 0.15\gamma_0$ and the same parameters as in Fig.~\ref{fig:fig6}(a). Approaching the $\bs{K}$-point from  $(\bs{K}-0.7 \hat{\bs{y}})$, states are occupied close to the chemical potential $\mu_t$. Starting from $eV = 0.15\gamma_0$ and increasing the voltage, first states close to $\bs{k}$-values  $ (\bs{K}-0.2 \hat{\bs{y}})$ contribute to the current and give rise to the first step-like increase of the conductance. 
Second states close the $(\bs{K}-0.2 \hat{\bs{K}})$ are occupied by increasing the voltage and the conductance gradually increases a second time. A similar argument holds for the suppression of the conductance close to voltages $eV = 0.4\gamma_0$ which is related to the resonant absorption of two photons.

In addition to a suppression of the differential conductance, the light-matter interaction induces a peak at $eV = 0.2\gamma_0$ [black arrow in Fig.~\ref{fig:fig6}(a)]. The process corresponding to the peak is sketch in Fig.~\ref{fig:fig6}(b) and is related to absorption of a single photon at energy $\mu_{\idxt}=0.2 \gamma_0$. Figure~\ref{fig:fig6}(b) shows the density of states summed over all momenta in the first Brillouin zone of the top and bottom layer. In the top layer, the states are occupied to the chemical potential $\mu_{\idxt}$. The light-matter interaction excites an electron to energies $\eps = \mu_{\idxt} +\omega$ matching the energy of the van Hove singularity of the top layer. The excited electron then tunnels to the van Hove singularity of the bottom layer providing a large density of unoccupied states. The peak at  $eV = 0.2\gamma_0$ in the conductance is hence a result of the overlap of the van Hove singularities of both the top and the bottom layer.

\subsection{Tunneling between graphene and graphene with finite mass term}
\begin{figure}[t!]			
	\includegraphics[width=0.9\linewidth,angle=0.]{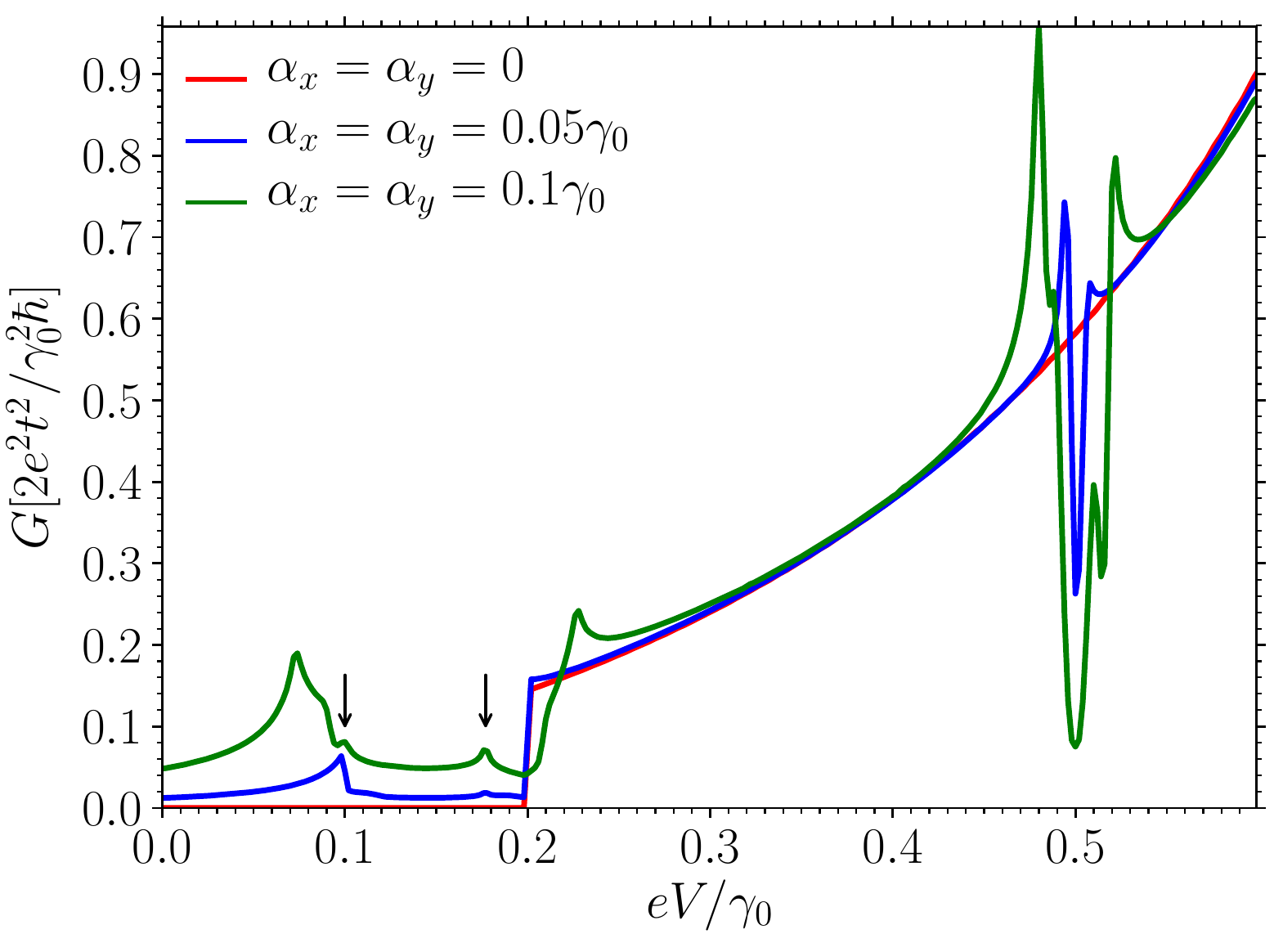} 
	\caption{Differential conductance for circular polarized light and different coupling strength $\alpha_x$ and $\alpha_y$. Compared to Fig.~\ref{fig:fig7}(a), the frequeny is $\omega=0.7\gamma_0$. The other parameters are $\omega = 0.8 \gamma_0$, $n=0,\pm1,\pm2$, $T=0$, $\eta= 0.001\gamma_0$. The Dirac point energy is set to $\eps_D=-0.2\gamma_0$ such that the differential conductance is suppressed close to voltages $eV \simeq \omega/2+\eps_D = 0.2\gamma_0$. The finite conductance in the gap depends on the $\eta=0.001\gamma_0$. The steps in the case of $\alpha_x=\alpha_y=0.1\gamma_0$ are a signature of the non-linear band structure of graphene for $\bs{k} \gg (\bs{K},\bs{K}^\prime)$. The two black arrows indicate the tunneling processes which are associated with the tunneling of photo-excited electrons to energies of the van Hove singularity of either the top or bottom layer.}   
	\label{fig:fig8}
\end{figure}

So far we discussed the differential conductance for the ideal case that light is absorbed in the top layer. Since the light is only partially absorbed in graphene~\cite{nair:2008}, also the lower layer will interact with the light. However, because of a substrate or interaction in graphene\cite{zhou:2007,Stroucken:2011}, a band gap will open. In this section, we assume that the bandgap of the bottom layer is larger than the frequency of the light such that the lower layer becomes essentially transparent. In a more general setting, the bottom layer can be made of any two-dimensional material with a bandgap larger than the frequency of the light. 
In Fig.~\ref{fig:fig7}(a) and \ref{fig:fig8}, we consider a finite mass $m_{\idxb}$ in the bottom layer such that the energy of the bare graphene Hamiltonian is given by $\varepsilon_{\bs{k}_{\idxb}} = - \gamma_0 \sqrt{\vert f(\bs{k}_{\idxb})\vert^2 + m_{\idxb}^2 } -(\mu_{\idxb}-\eps_D)$ and the bandgap becomes $2 m_{\idxb}$.

Figure~\ref{fig:fig7}(a) shows the differential conductance with the same parameters as in Fig.~\ref{fig:fig6}(a) except that now the bottom layer has a mass term $m_{\idxb}=0.4\gamma_0$ and the bandgap is $0.8\gamma_0$.
Without the light $\alpha_x=\alpha_y=0$, the current is completely suppressed at voltages $eV<m_{\idxb}$ because of the bandgap in the bottom layer. For $eV>0.2\gamma_0$, the voltage is larger than the bandgap on the bottom graphene and electrons can tunnel between the layers. We remark that the Dirac point energy is at $\eps_D = - 0.2\gamma_0$ such that the gap edge on the bottom layer is reached for voltages $eV = 0.2\gamma_0$. When the voltage is larger than the gap of the bottom layer, the behaviour is similar to the differential conductance in Fig.~\ref{fig:fig6}(a). 

As we have seen in previous sections, a finite light-matter interaction can occupy states above the chemical potential. We expect that the photoexcited electrons contribute to a current although the voltage is smaller than the gap on the bottom lead. This behavior is seen in Fig.~\ref{fig:fig7}(a) at finite light-matter coupling strength. The differential conductance for $eV<0.2\gamma_0$ is solely due to the tunnelling of photoexcited electrons. However, since the energies of electrons that are excited by a single photon are smaller than the bandgap edge at $eV=0.2\gamma_0$ of the bottom layer, the current is strongly suppressed for $eV<0.2\gamma_0$. In this case, the two-photon absorption is the dominant processes to the conductance.

Additionally to the finite conductance at voltages $eV < 0.2\gamma_0$, the differential conductance shows a peak at $eV \approx 0.28 \gamma_0$ indicated by the black arrow in Fig.~\ref{fig:fig7}(a). The process corresponding to such a peak is sketch in Fig.~\ref{fig:fig7}(b) and its origin is similar to the peak at $eV = 0.2\gamma_0$ in Fig.~\ref{fig:fig6}(a), namely electrons at the chemical potential are photoexcited and can tunnel into a large number of empty states due to the van Hove singularity of the bottom layer. The finite mass term in the bottom layer shifts the van Hove singularity to higher energies such that the peak in Fig.~\ref{fig:fig7}(a) appears at $eV \approx 0.28 \gamma_0$. In comparison to Fig.~\ref{fig:fig6}(b), the energies of the photoexcited electrons at energy $\eps \approx 0.88 \gamma_0$ are larger than the van Hove singularity of the top layer but match the energy of the van Hove singularity of the bottom layer.  

In Fig.~\ref{fig:fig8} we discuss the differential conductance at higher frequency $\omega=0.7\gamma_0$ and the same parameters as in Fig.~\ref{fig:fig7}(a). In this case, the conductance shows complex features even below the bandgap $eV<0.2\gamma_0$. Because of the high frequency $\omega=0.7\gamma_0$  and the opening of the gap in the quasienergy spectrum of graphene, the absorption of a single photon is sufficient to excite an electron above the bandgap edge of the bottom layer. The resonant absorption of a single photon gives then rise to the increase of the conductance at voltages $eV\approx 0.08\gamma_0$. 

Additionally, two smaller peaks occur at voltages $eV=0.1\gamma_0$ and $eV=0.18\gamma_0$ indicated by the black arrows in Fig.~\ref{fig:fig8}. The origin of both peaks can be associated with the van Hove singularities of both the top and the bottom layer. The van Hove singularity of the bottom layer occurs at energy $\eps=0.8\gamma_0$ due to the Dirac point energy $\eps_D=-0.2\gamma_0$. However, similar to Fig.~\ref{fig:fig7}(b), the van Hove singularity of the bottom layer appears at energies  $\eps\approx 0.88\gamma_0$ because of the finite mass term $m_\idxb$. When an electron at voltage $eV=0.1\gamma_0$ absorbs a photon, it is excited to energies of the van Hove singularity of the top layer and then tunnels to the bottom layer. The opposite process is also possible corresponding to first the excitation of an electron at voltages $eV=0.18\gamma_0$ and second, the tunneling to the energy of the van Hove singularity of the bottom layer.

%
%
%
%
\section{Conclusion}
\label{sec:conclusions}
We studied the transport properties of two graphene layers vertically coupled with an insulating layer. A voltage is applied between the top and the bottom layer giving rise to a tunnelling current. In such a contact, we discussed the differential conductance  when the top graphene layer is driven to a nonequilibrium state by an external light field. 

In agreement with previous result, multiple gaps open in the quasienergy spectrum of graphene depending on the polarization. We obtained that the gaps corresponding to a resonant absorptions/emission of a single photon between the conduction and valence band have different sizes due to the momentum dependence of the optical matrix element. This effect becomes in particular important at frequencies $\omega \gtrsim 0.5\gamma_0$ where the non-linearity of the bandstructure of graphene can not be neglected. 

We studied the differential conductance in two kinds of contacts. First, we considered the differential conductance between two graphene layers. Second, we studied the modification of the conductance when a gap opens in the bottom layer. In both cases, the effects of the light-matter interaction are manifested in a strong suppression of the conductance at voltages $eV =n\omega/2+\eps_D$ due to the opening of gaps in the quasienergy spectrum of graphene. In the case of strong driving, the different gap sizes at energies $\eps =n\omega/2+\eps_D $ become apparent in the differential conductance. Additional peaks in the differential conductance can be related to the tunneling of photo-excited electrons to the energies of the van Hove singularity of the top or bottom layer. For the case of a finite bandgap in the bottom layer, the differential conductance is solely due to the photo-excited electrons and is hence a direct signature of the light-matter interaction. Finally, we remark that although we studied the tunneling between two graphene layers, our approach can be extended to tunneling between arbitrary two-dimensional materials.

We conclude with neglected interaction and an outlook about extension to our approach. The light-matter interaction in graphene has been subject of much research and the nonequilibrium driving of graphene has been investigated in various approaches. Most of these approaches focus on the noninteracting picture in which the Coulomb interaction is neglected. However, it has been shown that the electron-electron interaction in two-dimensional materials can be tuned by engineering the surrounding dielectric environment. 
The electron-electron interaction has a significant effect on the band structure. For instance, Coulomb interaction leads to an open of a gap in the bandstructure and the formation of excitons \cite{Stroucken:2011,mueller:2018}.

The continuous applications of a laser pulse or light field to graphene results in heating and eventually damage of the sample. Under an intense laser pulse, the electron-phonon coupling in graphene will excite phonons and relax the electronic system. The phonon dissipation in the graphene under light irradiation and the effect on the transport properties can be studied by coupling the electronic system to a bath of phonons similar to the approaches in Refs.~\onlinecite{dehghani:2014,chen:2018v}.

\acknowledgments 
We acknowledge S.~Lara-Avila and H.~He for interesting discussions and comments. 
This work was supported by the Swedish Foundation for Strategic Research (SSF).

\appendix

\section{Keldysh Green's functions}
\label{app:GFandSE}
In the appendix, we recall the definitions of the Green's function and summarize the main steps to calculate the retarded, advanced and lesser Greens functions. We refer to the books of Refs.~[\onlinecite{Rammer:2007},\onlinecite{Cuevas-Scheer:2010}] for a detailed introduction.

Using the contour-ordered Green's function defined in Eq.~\eqref{eq:GFdefinition}, we derive the Dyson equation by using the Heisenberg equation of motion. We then transform the contour-ordered Green’s functions to the real-time and define the Green's function in Keldysh space as
\begin{equation}
G^{\lambda}_{\boldsymbol{k}\boldsymbol{k}^{}}(t,t^\prime) =
\begin{pmatrix}
G^{\lambda,11}_{\boldsymbol{k}\boldsymbol{k}^{}}(t,t^\prime)
&
G^{\lambda,12}_{\boldsymbol{k}\boldsymbol{k}^{}}(t,t^\prime)
\\[1mm]
G^{\lambda,21}_{\boldsymbol{k}\boldsymbol{k}^{}}(t,t^\prime)
&
G^{\lambda,22}_{\boldsymbol{k}\boldsymbol{k}^{}}(t,t^\prime)
\end{pmatrix} \, .
\label{eq:GFmatrix_app}
\end{equation}
In the above expression, the upper indexes $1$ or $2$ refer to the position of the times $t$ and $t^\prime$ on the Keldysh contour.
In addition to the Green's function in Eq.~\eqref{eq:GFmatrix}, we define the retarded Green's function $G^{\lambda,R}_{\boldsymbol{k}\boldsymbol{k}^{}}(t,t^\prime)$ and the advanced Green's function $G^{\lambda,A}_{\boldsymbol{k}\boldsymbol{k}^{}}(t,t^\prime)$.

These Greens functions and the ones in the Eq.~\eqref{eq:GFmatrix} are defined as
\begin{align}
{G}^{\lambda,11}_{\boldsymbol{k}\boldsymbol{k}^{\prime}}(t,t^{\prime})&=
-i \langle {\mathcal{T}} \hat{a}^{\lambda}_{\boldsymbol{k}}(t)\hat{a}_{\boldsymbol{k}^{\prime}}^{\lambda^{\dagger}}(t^{\prime}) \rangle 
\\
{G}^{\lambda,22}_{\boldsymbol{k}\boldsymbol{k}^{\prime}}(t,t^{\prime}) &=
-i \langle \tilde{\mathcal{T}} \hat{a}^{\lambda}_{\boldsymbol{k}}(t)\hat{a}_{\boldsymbol{k}^{\prime}}^{\lambda^{\dagger}}(t^{\prime}) \rangle   
\\
{G}^{\lambda,12}_{\boldsymbol{k}\boldsymbol{k}^{\prime}}(t,t^{\prime}) &=
i \langle \hat{a}_{\boldsymbol{k}^{\prime}}^{\lambda^{\dagger}}(t^\prime)  \hat{a}^{\lambda}_{\boldsymbol{k}^{}}(t^{})  \rangle  
\\
{G}^{\lambda,21}_{\boldsymbol{k}\boldsymbol{k}^{\prime}}(t,t^{\prime}) &=
-i\langle \hat{a}^{\lambda}_{\boldsymbol{k}}(t)  \hat{a}_{\boldsymbol{k}^{\prime}}^{\lambda^{\dagger}}(t^{\prime})  \rangle  
\\
{G}^{\lambda,R}_{\boldsymbol{k}\boldsymbol{k}^{\prime}}(t,t^{\prime}) &=
-i \theta(t-t^{\prime})\langle \{ \hat{a}^{\lambda}_{\boldsymbol{k}}(t) ,  \hat{a}_{\boldsymbol{k}^{\prime}}^{\lambda^{\dagger}}(t^{\prime}) \} \rangle  
\\
{G}^{\lambda,A}_{\boldsymbol{k}\boldsymbol{k}^{\prime}}(t,t^{\prime}) &=
i \theta(t^{\prime}-t)\langle \{\hat{a}^{\lambda}_{\boldsymbol{k}}(t) , \hat{a}_{\boldsymbol{k}^{\prime}}^{\lambda^{\dagger}}(t^{\prime}) \}  \rangle  
\end{align}

The real time-ordering and anti-time ordering operators are denoted by $\mathcal{T}$ and $\tilde{\mathcal{T}}$, respectively. 
The anti-commutator is denoted by $\{\,,\,\}$.

The Green's functions satisfy the following relations which are useful to derive the retarded and advanced Green's functions. The Green's functions are related by
\begin{align}
G_{\boldsymbol{k}\boldsymbol{k}^{\prime}}^{\lambda,R}(t,t^{\prime})-G_{\boldsymbol{k}\boldsymbol{k}^{\prime}}^{\lambda,A}(t,t^{\prime})&=G_{\boldsymbol{k}\boldsymbol{k}^{\prime}}^{\lambda,>}(t,t^{\prime})-G_{\boldsymbol{k}\boldsymbol{k}^{\prime}}^{\lambda,<}(t,t^{\prime}) \\
G_{\boldsymbol{k}\boldsymbol{k}^{\prime}}^{\lambda,K}(t,t^{\prime})&=G_{\boldsymbol{k}\boldsymbol{k}^{\prime}}^{\lambda,11}(t,t^{\prime}) + {G}_{\boldsymbol{k}\boldsymbol{k}^{\prime}}^{\lambda,22}(t,t^{\prime}) \nonumber \\ &= G_{\boldsymbol{k}\boldsymbol{k}^{\prime}}^{\lambda,<}(t,t^{\prime}) + {G}_{\boldsymbol{k}\boldsymbol{k}^{\prime}}^{\lambda,>}(t,t^{\prime})  \\
G_{\boldsymbol{k}\boldsymbol{k}^{\prime}}^{\lambda,R}(t,t^{\prime})&=G_{\boldsymbol{k}\boldsymbol{k}^{\prime}}^{\lambda,11}(t,t^{\prime})-G_{\boldsymbol{k}\boldsymbol{k}^{\prime}}^{\lambda,<}(t,t^{\prime}) \nonumber \\ &=G_{\boldsymbol{k}\boldsymbol{k}^{\prime}}^{\lambda,>}(t,t^{\prime})-{G}_{\boldsymbol{k}\boldsymbol{k}^{\prime}}^{\lambda,22}(t,t^{\prime}) \\
G_{\boldsymbol{k}\boldsymbol{k}^{\prime}}^{\lambda,A}(t,t^{\prime})&=G_{\boldsymbol{k}\boldsymbol{k}^{\prime}}^{\lambda,<}(t,t^{\prime})-{G}_{\boldsymbol{k}\boldsymbol{k}^{\prime}}^{\lambda,22}(t,t^{\prime}) \nonumber \\ &=G_{\boldsymbol{k}\boldsymbol{k}^{\prime}}^{\lambda,11}(t,t^{\prime})-{G}_{\boldsymbol{k}\boldsymbol{k}^{\prime}}^{\lambda,>}(t,t^{\prime}) \\
G_{\boldsymbol{k}\boldsymbol{k}^{\prime}}^{\lambda,11}(t,t^\prime)&=G_{\boldsymbol{k}\boldsymbol{k}^{\prime}}^{\lambda,R}(t,t^\prime)+G_{\boldsymbol{k}\boldsymbol{k}^{\prime}}^{\lambda,<}(t,t^\prime) \nonumber \\ &= G_{\boldsymbol{k}\boldsymbol{k}^{\prime}}^{\lambda,A}(t,t^\prime)+G_{\boldsymbol{k}\boldsymbol{k}^{\prime}}^{\lambda,>}(t,t^\prime) \\
G_{\boldsymbol{k}\boldsymbol{k}^{\prime}}^{\lambda,22}(t,t^\prime)&=G_{\boldsymbol{k}\boldsymbol{k}^{\prime}}^{\lambda,<}(t,t^\prime)-G_{\boldsymbol{k}\boldsymbol{k}^{\prime}}^{\lambda,A}(t,t^\prime) \nonumber \\ &= G_{\boldsymbol{k}\boldsymbol{k}^{\prime}}^{\lambda,>}(t,t^\prime)-G_{\boldsymbol{k}\boldsymbol{k}^{\prime}}^{\lambda,R}(t,t^\prime)
\end{align}
and
\begin{align}
G_{\boldsymbol{k}\boldsymbol{k}^{\prime}}^{\lambda,<}(t,t^{\prime})&= (G_{\boldsymbol{k}\boldsymbol{k}^{\prime}}^{\lambda,K}(t,t^{\prime})-G_{\boldsymbol{k}\boldsymbol{k}^{\prime}}^{\lambda,R}(t,t^{\prime})+G_{\boldsymbol{k}\boldsymbol{k}^{\prime}}^{\lambda,A}(t,t^{\prime}))/2 \\
G_{\boldsymbol{k}\boldsymbol{k}^{\prime}}^{\lambda,>}(t,t^{\prime})&= (G_{\boldsymbol{k}\boldsymbol{k}^{\prime}}^{\lambda,K}(t,t^{\prime})+G_{\boldsymbol{k}\boldsymbol{k}^{\prime}}^{\lambda,R}(t,t^{\prime})-G_{\boldsymbol{k}\boldsymbol{k}^{\prime}}^{\lambda,A}(t,t^{\prime}))/2  
\end{align}
Further, the hermitian-conjugate of the electron Green's functions satisfy the relations
\begin{align}
{G_{\boldsymbol{k}\boldsymbol{k}^{\prime}}^{\lambda,R}(t,t^\prime)}^* &= G_{\boldsymbol{k}^{\prime}\boldsymbol{k}}^{\lambda,A}(t^\prime,t) \\
{G_{\boldsymbol{k}\boldsymbol{k}^{\prime}}^{\lambda,<}(t,t^\prime)}^* &= - G_{\boldsymbol{k}^{\prime}\boldsymbol{k}}^{\lambda,<}(t^\prime,t) \\
{G_{\boldsymbol{k}\boldsymbol{k}^{\prime}}^{\lambda,>}(t,t^\prime)}^* &= - G_{\boldsymbol{k}^{\prime}\boldsymbol{k}}^{\lambda,>}(t^\prime,t) \\
{G_{\boldsymbol{k}\boldsymbol{k}^{\prime}}^{\lambda,11}(t,t^\prime)}^*& = -G_{\boldsymbol{k}^{\prime}\boldsymbol{k}}^{\lambda,22}(t^\prime,t) \\
{G_{\boldsymbol{k}\boldsymbol{k}^{\prime}}^{\lambda,22}(t,t^\prime)}^* &= -G_{\boldsymbol{k}^{\prime}\boldsymbol{k}}^{\lambda,11}(t^\prime,t)
\end{align}
\section{Recursive solution of Green's functions}
\label{app:RecursiveGFSolution}
As discussed in Sec.~\ref{sec:method}, the retarded, advanced and lesser Green's functions are the basic building blocks to calculate the dc-current in Eq.~\eqref{eq:current}. In this appendix, we describe the recursive method to compute the retarded Green's functions $\hat{G}_{\bs{k}_{},nm}^{R}(\eps)$ in Eq.~\eqref{eq:GR_left} following Refs.~\onlinecite{cuevas:1996,cuevas:2001}. The advanced Green's function is related to the retarded Green's function by $\hat{G}_{\bs{k}_{},nm}^{A}(\eps)=\hat{G}_{\bs{k}_{},mn}^{R^\dagger}(\eps) $  and the lesser Green's function is obtained from Eq.~\eqref{eq:G12_left}. 

From the relation of the Green's functions
$
\hat{G}^{R}_{\boldsymbol{k},nm}(\varepsilon) =
\hat{G}^{R,12}_{\boldsymbol{k},n-m}(\varepsilon+(n+m)\omega/2)
$, one can show that
$
\hat{G}^{R}_{\boldsymbol{k},nm}(\varepsilon) =  \hat{G}^{R}_{\boldsymbol{k},n-m0}(\varepsilon+m\omega) 
$. 
In the following it is hence sufficient to set $m=0$ and define the transfer matrices $ \hat{z}^{\pm}_n$ by (neglecting the energy and momentum dependence)
\begin{equation}
\hat{G}^{R}_{n0} = \hat{z}^{\pm}_n \hat{G}^{R}_{n\mp 10} \, ,
\label{eq:apptransfermatrix}
\end{equation}
where $ \hat{z}^{+}_n$ and $ \hat{z}^{-}_n$ are defined  for $n>1$ and  $n<-1$, respectively. 
Using the recursive equation of the retarded Green's function [Eq.~\eqref{eq:GR_left}]
\begin{align}
\hat{G}^{R}_{n0} = \hat{g}^{R}_{n} \delta_{n0} 
-
{\hat{g}^{R}_{n}}  \hat{M}^{+}_{} \hat{G}^{R}_{n+10} -
{\hat{g}^{R}_{n}} \hat{M}^{-}_{} \hat{G}^{R}_{n-10} \, ,
\label{eq:appgRdyson}
\end{align}
it follows that the transfer matrices satisfy the relation

\begin{equation}
 \hat{z}^{\pm}_n  = -(\hat{g}_n^{R^{-1}}+M^{\pm}  \hat{z}^{\pm}_{n+1})^{-1} M^{\mp} \, .
 \label{eq:appzn}
\end{equation}

The recursive method then works as follows. Starting from large $n$ we set the elements $ \hat{z}^{\pm}_{n+1}$ and $ \hat{z}^{\pm}_{-n-1}$ to zero, and calculate the remaining transfer matrices  
$ \hat{z}^{+}_{n}, \dots , \hat{z}^{+}_{1}$ and $ \hat{z}^{-}_{-n}, \dots , \hat{z}^{-}_{-1}$ by Eq.~\eqref{eq:appzn}. 
Setting $n=0$ in Eq.~\eqref{eq:appgRdyson}  and using Eq.~\eqref{eq:apptransfermatrix}, the Green's function $\hat{G}_{00}^R$ is given by 
\begin{equation}
\hat{G}_{00}^R = (\hat{g}_{0}^{R^{-1}} + M^+  \hat{z}_{1}^+ +M^- \hat{z}_{-1}^{-})^{-1} \, . 
\end{equation} 
To obtain the Green's functions for arbitrary $n$, we use the solution of $\hat{G}_{00}^R $ and apply Eq.~\eqref{eq:apptransfermatrix}.

\bibliographystyle{apsrev-titles}
\bibliography{references}

\end{document}